\documentclass[aps,prr,reprint,superscriptaddress]{revtex4-2}

\usepackage{amsmath}
\usepackage{babel}
\usepackage{graphicx}
\graphicspath{ {figs/} }
\usepackage{xcolor}
\usepackage{longtable}
\usepackage{standalone}
\usepackage{tikz}
\usetikzlibrary{arrows.meta,calc,decorations.pathreplacing,calligraphy}

\begin{document}


\title{Extracting conformal data from finite-size tensor-network flow in critical two-dimensional classical models}

\author{Sing-Hong Chan}
\affiliation{ Department of Physics, National Tsing Hua University, Hsinchu 30013, Taiwan}

\author{Pochung Chen}
\email{pcchen@phys.nthu.edu.tw}
\affiliation{ Department of Physics, National Tsing Hua University, Hsinchu 30013, Taiwan}
\affiliation{ Physics Division, National Center for Theoretical Sciences, Taipei 10617, Taiwan}
\affiliation{ Center for Theory-Computation-Data Science Research, National Tsing Hua University, Hsinchu 30013, Taiwan}




\date{\today}

\begin{abstract}
  We present a general framework for extracting conformal data from critical two-dimensional classical lattice models using finite-size tensor-network flow.
  The central idea is to identify, from transfer-matrix spectra, a self-consistent finite-size window 
  together with a crossover scale that separates the finite-size-scaling regime from the finite-entanglement-scaling regime induced by bond-dimension truncation.
  Within this window, the central charge, scaling dimensions, and conformal spins can be estimated 
  without requiring a unique critical fixed-point tensor or detailed prior knowledge of the underlying conformal field theory.
  We benchmark the framework using three tensor-network renormalization schemes for the critical two-dimensional Ising and three-state clock models.
  Across schemes, we find robust universal behavior below the crossover scale, enabling accurate extraction of conformal data up to relatively high conformal levels.
  The analysis also yields a natural operational definition of entanglement scaling for classical tensor-network calculations and, in turn, a complementary estimator of the central charge.
\end{abstract}


\maketitle


\section{Introduction}

Tensor-network methods provide a natural real-space framework for studying classical lattice models, with direct connections to renormalization-group ideas, critical phenomena, and entanglement structure.
By representing the partition function as a tensor network, one can evaluate thermodynamic quantities accurately at system sizes far beyond the reach of exact diagonalization \cite{Levin:2007ju}.
At continuous phase transitions, two-dimensional classical lattice models are described in the continuum limit by conformal field theory (CFT) \cite{Cardy:1984ue, Cardy.1986, Blote:1986ui}.
An important challenge is therefore to extract conformal data, such as central charge, scaling dimensions, and conformal spins, directly from lattice calculations.

A natural route is to use tensor-network renormalization to access critical fixed-point tensors and extract conformal data from their structure \cite{Li.2022ybe}.
This strategy has proved powerful, but it is also technically delicate.
Short-range entanglement structures such as corner double-line contributions and gauge redundancies must be handled carefully \cite{Gu.2009, Lyu.2021, Li.2022ybe, Guo.2024}.
More fundamentally, while fixed-point-tensor methods work well in practice, 
a rigorous and fully general definition of tensor-network renormalization-group fixed points remains an active subject of investigation \cite{Kennedy.2022, Kennedy.2024, Ebel.20251iu}.
These considerations motivate approaches to conformal-data extraction that do not rely on the existence of a unique, well-behaved critical fixed-point tensor.

Here we formulate conformal-data extraction as a problem in finite-size tensor-network flow.
Although tensor-network methods are naturally suited to real-space renormalization, finite-size scaling in tensor-network calculations has been explored less extensively.
Recent work has combined tensor-network methods with finite-size scaling analyses, including data-collapse and crossing-point techniques, 
to determine critical points and exponents in deformed Affleck-Kennedy-Lieb-Tasaki states and two-dimensional classical lattice models \cite{Huang.2020wh, Huang.2023, Maiti.2025}.
Related studies of Fisher zeros and Lee-Yang zeros have also pointed to the effectiveness of tensor-network-enabled finite-size analyses \cite{Hong:2019cm, Hong.2022}.
These developments suggest that finite-size tensor-network flow can provide a general route to universal conformal information.

In this work we develop such a framework for critical two-dimensional classical lattice models.
The central ingredients are transfer-matrix spectra on finite-width strips, 
a self-consistent finite-size window identified using numerically estimated conformal spin, 
and a crossover scale $L_x^*$ that marks the onset of finite-entanglement effects induced by bond-dimension truncation.
Conformal data evaluated at $L_x^*$ are taken as the optimal estimates for a given set of control parameters.
This makes it possible to extract central charge, scaling dimensions, and conformal spins using only general structural properties of CFT, 
without requiring a unique critical fixed-point tensor or detailed prior knowledge of the underlying CFT.

The framework is compatible with multiple tensor-network renormalization schemes.
We benchmark it using higher-order tensor renormalization group (HOTRG) \cite{Xie:2012iy}, periodic transfer-matrix renormalization group (PTMRG) \cite{Fedorovich.2025}, and core-tensor renormalization group (CTRG) \cite{Lan:2019et}, and apply it to the critical two-dimensional Ising and three-state clock models, where exact CFT data are known.
Across methods, we find robust universal behavior below the crossover scale, while the crossover itself depends on the observable, the bond dimension, and the renormalization scheme.
For the models studied here, HOTRG provides the best overall accuracy for conformal-data extraction.

Our main result is that conformal data can be extracted accurately from finite-size tensor-network flow up to relatively high conformal levels.
The framework also yields a natural operational definition of entanglement scaling for classical systems 
and thereby provides a complementary estimator of the central charge in addition to the conventional ground-state-energy estimator.

The rest of this paper is organized as follows.
Section~\ref{sec:TM} reviews the transfer-matrix formalism and finite-size estimators of conformal data.
Section~\ref{sec:TN} summarizes the tensor-network renormalization schemes and transfer-matrix construction.
Section~\ref{sec:results} presents the main numerical results, with detailed model-by-model analyses in Secs.~\ref{sec:Ising} and \ref{sec:3-clock}.
Two complementary central-charge estimators are discussed in Sec.~\ref{sec:c}, followed by conclusions and outlook.

\section{Transfer matrix formalism \label{sec:TM}}

Consider a two-dimensional classical system on an $L_x \times L_y$ torus with periodic boundary conditions.
Its partition function can be written as
\begin{equation}
  Z(L_x,L_y)=\mathrm{Tr}\left[\mathcal{T}^{L_y}(L_x)\right],
\end{equation}
where $\mathcal{T}(L_x)$ is the row-to-row transfer matrix.
At criticality, the low-energy spectrum of the transfer matrix on an infinitely long strip is governed by the underlying CFT \cite{Cardy:1984ue, Cardy.1986, Blote:1986ui}.
Writing
\begin{equation}
  \mathcal{T}(L_x)=e^{-H(L_x)},
\end{equation}
we interpret $H(L_x)$ as the corresponding effective lattice Hamiltonian.
Let
\begin{equation}
  H(L_x)\,|E_i(L_x)\rangle=E_i(L_x)\,|E_i(L_x)\rangle,
\end{equation}
then equivalently
\begin{equation}
  \mathcal{T}(L_x)\,|E_i(L_x)\rangle=e^{-E_i(L_x)}|E_i(L_x)\rangle.
\end{equation}
Here we assume $i=0,1,2,\ldots$ and $E_0\le E_1\le E_2\le\cdots$.
By state-operator correspondence, we expect that each level $|E_i\rangle$ is associated with a CFT operator $\phi_i$ \cite{Cardy:1984ue, Cardy.1986, Blote:1986ui}.

For large $L_x$, finite-size scaling gives
\begin{equation}
  \frac{E_i(L_x)}{L_x}=f_\infty+\frac{2\pi}{L_x^2}\left(\Delta_{\phi_i}-\frac{c}{12}\right)+\mathcal{O}(\cdots),
  \label{eq:FSS_Ei}
\end{equation}
where $\Delta_{\phi_i}$ is the scaling dimension associated with $\phi_i$, $c$ is the central charge, $f_\infty$ is the bulk free-energy density, and $\mathcal{O}(\cdots)$ denotes subleading corrections.
For unitary CFTs, the ground state corresponds to the identity operator with $\Delta_{\mathbf{I}}=0$, so
\begin{equation}
  \frac{E_0(L_x)}{L_x}=f_\infty-\frac{\pi c}{12L_x^2}+\mathcal{O}(\cdots).
\end{equation}
This motivates the estimator
\begin{equation}
  c_{E_0}(L_x)\equiv-\frac{6}{\pi}\frac{d}{dL_x^2}\left[\frac{E_0(L_x)}{L_x}\right]\xrightarrow[L_x\to\infty]{}c,
  \label{eq:c_E0}
\end{equation}
where the subscript $E_0$ distinguishes it from the entropy-based estimator introduced below.

Similarly, the excitation gap yields scaling dimensions:
\begin{equation}
  \frac{E_i(L_x)-E_0(L_x)}{L_x}=\frac{2\pi}{L_x^2}\Delta_{\phi_i}+\mathcal{O}(\cdots),
\end{equation}
and we define
\begin{equation}
  X_i(L_x)\equiv\frac{[E_i(L_x)-E_0(L_x)]L_x}{2\pi}\xrightarrow[L_x\to\infty]{}\Delta_{\phi_i}.
\end{equation}

The conformal spin is extracted from lattice momentum.
Let $\mathcal{T}_P$ be the one-site translation operator along the $x$ direction.
Since $[\mathcal{T},\mathcal{T}_P]=0=[H,\mathcal{T}_P]$, the eigenstates can be chosen simultaneously:
\begin{equation}
  \mathcal{T}_P|E_i(L_x),S_i(L_x)\rangle
  =e^{i\frac{2\pi}{L_x}S_i(L_x)}|E_i(L_x),S_i(L_x)\rangle,
\end{equation}
so $S_i(L_x)$ converges to the CFT conformal spin $S_{\phi_i}$.
In practice, we diagonalize $\mathcal{T}\mathcal{T}_P$:
\begin{equation}
  \mathcal{T}\mathcal{T}_P|E_i(L_x),S_i(L_x)\rangle
  =e^{-E_i(L_x)+i\frac{2\pi}{L_x}S_i(L_x)}|E_i(L_x),S_i(L_x)\rangle.
\end{equation}
The real and imaginary parts of the eigenvalues provide $E_i(L_x)$ and $\frac{2\pi}{L_x}S_i(L_x)$, respectively.

Finally, we define an entropy-based estimator for the central charge.
Consider an infinite strip of width $2L_x$ and the dominant eigenvector $|E_0(2L_x)\rangle$ of $\mathcal{T}(2L_x)$.
Viewing this state as the ground state of the quantum Hamiltonian $H(2L_x)$ of $2L_x$ sites, 
we bipartition the strip into two halves of width $L_x$ and compute the bipartite entanglement entropy $S_E(L_x,L_x)$.
At criticality we expect that,
\begin{equation}
  S_E(L_x,L_x)=\frac{c}{3}\ln\left(\frac{2 L_x}{\pi}\right)+\alpha,
\end{equation}
with non-universal constant $\alpha$.
Therefore we define
\begin{equation}
  c_S(L_x)\equiv3\frac{dS_E(L_x,L_x)}{d\ln L_x}\xrightarrow[L_x\to\infty]{}c,
  \label{eq:c_S}
\end{equation}
where the subscript $S$ indicates that the estimation is based on the entropy.

\section{Tensor network methods \label{sec:TN}}

In this section, we briefly review how we approximately construct the transfer matrix and the shift operator within tensor-network renormalization.
We employ three complementary tensor-network renormalization schemes: PTMRG, CTRG, and HOTRG, 
in order to cross-check universal quantities against method-dependent truncation effects.
Within our computational budget, HOTRG gives the best overall conformal-tower accuracy.
PTMRG and CTRG are used to provide consistency checks and to initialize larger block tensors for HOTRG.

Having defined finite-size estimators for the conformal data, we now explain how to evaluate them for system sizes beyond the reach of exact diagonalization.
Our strategy is to use tensor-network renormalization to construct approximations to the transfer matrix $\mathcal{T}$ and the shift operator $\mathcal{T}_P$.
We start from the tensor-network representation of the partition function
\begin{equation}
  Z=\mathrm{tTr}  \prod_{\mathrm{sites}}  \mathbf{T}_{ijkl}^{1\times 1},
\end{equation}
where the rank-4 tensor $ \mathbf{T}_{ijkl}^{1\times 1}$ represents a $1\times 1$ block.
This tensor is constructed by the standard procedure \cite{Huang.2023, Maiti.2025}.
We then apply tensor-network renormalization to obtain renormalized tensors $ \mathbf{T}_{ijkl}^{L_x \times L_y}$ that approximately represent $L_x \times L_y$ blocks.
Specifically, we implement the core-tensor renormalization group (CTRG) \cite{Lan:2019et}, the periodic transfer-matrix renormalization group (PTMRG) \cite{Fedorovich.2025}, and the higher-order tensor renormalization group (HOTRG) \cite{Xie:2012iy}.
Below we sketch the RG steps of each method and refer to the cited works for implementation details.

For PTMRG, we first initialize the horizontal, vertical, and bulk tensors to the same $1\times 1$ tensor: $\mathbf{T}_h^{1,1}=\mathbf{T}_v^{1,1}=\mathbf{T}^{1,1}$.
We then perform the RG in $x$-direction for the bulk and horizontal tensors
\begin{eqnarray}
  \begin{cases}
  & \text{PTMRG}_x \left[ \mathbf{T}^{L_x, L_y}, \mathbf{T}_v^{1, L_y} \right] \rightarrow \mathbf{T}^{L_x+1, L_y} \\
  & \text{PTMRG}_x \left[ \mathbf{T}_h^{L_x, 1}, \mathbf{T}^{1,1} \right] \rightarrow \mathbf{T}_h^{L_x+1,1}
   \end{cases},
\end{eqnarray}
followed by the RG in $y$-direction for the bulk and the vertical tensors
\begin{eqnarray}
  \begin{cases}
  & \text{PTMRG}_y \left[ \mathbf{T}^{L_x+1, L_y}, \mathbf{T}_h^{L_x+1, 1} \right] \rightarrow \mathbf{T}^{L_x+1, L_y+1} \\
  & \text{PTMRG}_y \left[ \mathbf{T}_v^{1, L_y}, \mathbf{T}^{1,1} \right] \rightarrow \mathbf{T}_v^{1, L_y+1} 
   \end{cases}.
\end{eqnarray}
After one complete RG, the effective size increases by one in both directions.

For CTRG, we first initialize the horizontal, vertical, core, and bulk tensors with the same $1\times 1$ tensor,
$\mathbf{T}_h^{1,1}=\mathbf{T}_v^{1,1}=\mathbf{T}_c^{1,1}=\mathbf{T}^{1,1}$.
We then decompose the core tensor into left and right pieces, $\mathbf{T}_c^{1,1}=\mathbf{T}_{cl}^{1,1} \mathbf{T}_{cr}^{1,1}$.
For the left core tensor, we perform the upper-left core-tensor RG step
\begin{equation}
  \text{CTRG} \left[ \mathbf{T}_h^{L_x, 1}, \mathbf{T}_{cl}^{1, 1}, \mathbf{T}_{cl}^{L_x, L_y} \right] \rightarrow \mathbf{T}_{cl}^{L_x+1, L_y+1},
\end{equation}
and similarly for the right core tensor,
\begin{equation}
  \text{CTRG} \left[ \mathbf{T}_v^{1, L_y}, \mathbf{T}_{cr}^{1, 1}, \mathbf{T}_{cr}^{L_x, L_y} \right] \rightarrow \mathbf{T}_{cr}^{L_x+1, L_y+1}.
\end{equation}
The left and right core tensors are then merged to obtain the renormalized core tensor,
\begin{equation}
  \text{Merge} \left[ \mathbf{T}_{cl}^{L_x+1, L_y+1}, \mathbf{T}_{cr}^{L_x+1, L_y+1} \right] \rightarrow \mathbf{T}_{c}^{L_x+1,L_y+1}.
\end{equation}
At the same time, we perform RG steps in the $x$ and $y$ directions for the horizontal and vertical tensors, respectively:
\begin{eqnarray}
  \begin{cases}
  & \text{CTRG}_x \left[ \mathbf{T}_h^{L_x, 1}, \mathbf{T}^{1,1} \right] \rightarrow \mathbf{T}_h^{L_x+1,1}  \\
  & \text{CTRG}_y \left[ \mathbf{T}_v^{1, L_y}, \mathbf{T}^{1,1} \right] \rightarrow \mathbf{T}_v^{1, L_y+1},
   \end{cases}
\end{eqnarray}
which are used in the next iteration.
After one complete RG step, the effective size of the core tensor increases by one in both directions.

For HOTRG, we first initialize an $L_0 \times L_0$ block tensor $\mathbf{T}^{L_x=L_0, L_y=L_0}$, where $L_0=2,3,5,7,9$.
For small $L_0$, we construct $\mathbf{T}^{L_0, L_0}$ by exact contraction of $\mathbf{T}_{ijkl}^{1\times 1}$.
For larger $L_0$, we use PTMRG with a high bond dimension to obtain an approximate $\mathbf{T}^{L_0, L_0}$.
We have verified that, for this range of $L_0$, the approximation is highly accurate.
We then perform the RG step in the $x$ direction,
\begin{equation}
  \text{HOTRG}_x \left[ \mathbf{T}^{L_x, L_y}, \mathbf{T}^{L_x, L_y} \right] \rightarrow \mathbf{T}^{2L_x, L_y},
\end{equation}
followed by the RG step in the $y$ direction,
\begin{equation}
  \text{HOTRG}_y \left[ \mathbf{T}^{2L_x, L_y}, \mathbf{T}^{2L_x, L_y} \right] \rightarrow \mathbf{T}^{2L_x, 2L_y}.
\end{equation}
After one complete RG step, the effective size doubles in both directions.
Starting from an $L_0 \times L_0$ block also gives access to a larger set of distinct sizes.
This added flexibility will be important for identifying the crossover length scale discussed later.

The shift operator can be constructed from the $1\times 1$ block shift operator $\mathbf{T}^{1\times 1}_P$ \cite{Hauru:2016cv}.
Whenever we perform an RG step in the $x$ direction, we use the same isometry to renormalize $\mathbf{T}^{1\times L_y}_P$.
For CTRG and PTMRG, one has
\begin{equation}
  \text{RG}_x \left( \mathbf{T}_P^{L_x \times 1} \right) \rightarrow \mathbf{T}_P^{ (L_x+1) \times 1 },
\end{equation}
while for HOTRG the block size doubles:
\begin{equation}
  \text{RG}_x \left( \mathbf{T}_P^{L_x \times 1} \right) \rightarrow \mathbf{T}_P^{ (2 L_2) \times 1 }.
\end{equation}

For all three schemes, it is necessary to set an upper bound $D$ for the bond dimension of the renormalized tensors.
The computational complexity typically scales as a power of $D$.
Specifically, HOTRG/PTMRG/CTRG scales as $\mathcal{O}(D^7)$, $\mathcal{O}(D^5)$, and $\mathcal{O}(D^4)$ respectively.
However, it takes $\mathcal{O}(L_y)$ RG steps for PTMRG/CTRG to reach size $L_y$, while HOTRG only takes $\mathcal{O}(\log L_y)$ steps.
Moreover, while it is possible to perform PTMRG/CTRG with larger $D$, 
it is also expected that a larger $D$ is needed to reach the same accuracy as HOTRG.
A priori, it is not clear which scheme can provide the best accuracy for a given computational budget.
As it will be shown later, our results show that HOTRG provides the best accuracy for all conformal data
for the models studied in this work.

After reaching the desired size $(L_x, L_y)$, we stack $n$ renormalized tensor $\mathbf{T}^{L_x \times L_y}$ horizontally,
and then trace out the index in $x$-direction. This results in a transfer matrix $ \widetilde{ \mathcal{T}}_{nL_x, L_y}(D)$,
which approximates the row-to-row transfer matrix with width $n L_x$, to the power of $L_y$.
\begin{equation}
   \widetilde{ \mathcal{T}}_{nL_x \times L_y}(D) 
   \equiv \text{Tr}_x \left( \mathbf{T}^{L_x \times L_y} \cdots  \mathbf{T}^{L_x \times L_y}  \right)
   \approx  \mathcal{T}^{L_y}(n L_x).
\end{equation}
In Fig.~\ref{fig:transfer} we sketch the construction of the row-to-row transfer matrix for the case of $n=3$.
On the other hand, by stacking $n$ renormalized shift operator $\mathbf{T}_P^{L_x \times 1}$
one obtains the  renormalized one-site shift matrix with width $n L_x$.
\begin{equation}
  \widetilde{ \mathcal{T}}_{P, nL_x \times 1} (D) 
   \equiv \text{Tr}_y \left( \mathbf{T}_P^{L_x \times 1} \cdots  \mathbf{T}_P^{L_x \times 1}  \right) \approx  \mathcal{T}_P(n L_x) .
\end{equation}
Finally we diagonalize the product of $ \widetilde{ \mathcal{T}}_{nL_x \times L_y}(D) $ and $ \widetilde{ \mathcal{T}}_{P, nL_x \times 1} (D) $.
From the real and  imaginary part of the eigenvalues, we obtain the energy $E_i(D, n L_x)$ and conformal spin $S_i(D, n L_x)$ respectively.
We note in passing that we enforce $\mathcal{Z}_2$ symmetry in the Ising model and $\mathcal{Z}_3$ symmetry in the 3-state clock model in our tensor-network calculations,
and the eigenvalue and eigenvector can be labeled with charge $Q$. 

\begin{figure}[t]
\begin{tikzpicture}[
    >={Stealth[length=2mm,width=1.6mm]},
    tens/.style={draw=black, line width=0.5pt, fill=blue!12, rounded corners=0.5pt,
                 minimum width=8mm, minimum height=8mm, inner sep=1pt,
                 font=\small},
    big/.style={draw=black, line width=0.5pt, fill=orange!18, rounded corners=0.5pt,
                minimum width=20mm, minimum height=10mm, inner sep=1pt,
                font=\scriptsize},
    leg/.style={line width=0.5pt},
    lg/.style={font=\scriptsize, inner sep=1.2pt}
  ]

  \foreach \i in {1,2,3}{
    \node[tens] (T\i) at (1.6*\i,0) {$\mathbf{T}$};
    \draw[leg] (T\i.north) -- ++(0,0.55) node[lg,above,inner sep=0.8pt] {$D$};
    \draw[leg] (T\i.south) -- ++(0,-0.55) node[lg,below,inner sep=0.8pt] {$D$};
  }

  \draw[leg] (T1) -- node[lg,above,inner sep=0.8pt]{$D$} (T2);
  \draw[leg] (T2) -- node[lg,above,inner sep=0.8pt]{$D$} (T3);

  \draw[leg,rounded corners=4pt]
        (T1.west) -- (0.9,0) -- (0.9,-1.35) -- (5.5,-1.35) -- (5.5,0) -- (T3.east);
  \node[lg,fill=white,inner sep=1.2pt] at (3.2,-1.35) {$\mathrm{Tr}_x$};

  \draw[decorate,decoration={calligraphic brace,amplitude=3pt,raise=1.5pt},
        line width=0.5pt]
        (0.9,1.2) -- (5.5,1.2) node[midway,above=5pt,font=\scriptsize] {$n=3$};

  \draw[->,line width=0.8pt] (5.95,0) -- (6.75,0);

  \node[big] (S) at (7.95,0) {$\widetilde{ \mathcal{T}}_{\,nL_x\times L_y}(D)$};
  \draw[leg] (S.north) -- ++(0,0.55) node[lg,above,inner sep=0.8pt] {$D^{\,n}$};
  \draw[leg] (S.south) -- ++(0,-0.55) node[lg,below,inner sep=0.8pt] {$D^{\,n}$};

\end{tikzpicture}
  \caption{Three copies of the rank-4 tensor $\mathbf{T}$ for an $L_x\times L_y$
  block, contracted along $x$ and closed by $\mathrm{Tr}_x$, yielding the
  row-to-row transfer matrix $\widetilde{ \mathcal{T}}_{nL_x\times L_y}(D)$ with vertical
  bond dimension $D^{n}$.}
  \label{fig:transfer}
\end{figure}

\section{Numerical results \label{sec:results}}

\begin{figure}[t]
  \begin{center}
    \includegraphics[width=0.95\columnwidth]{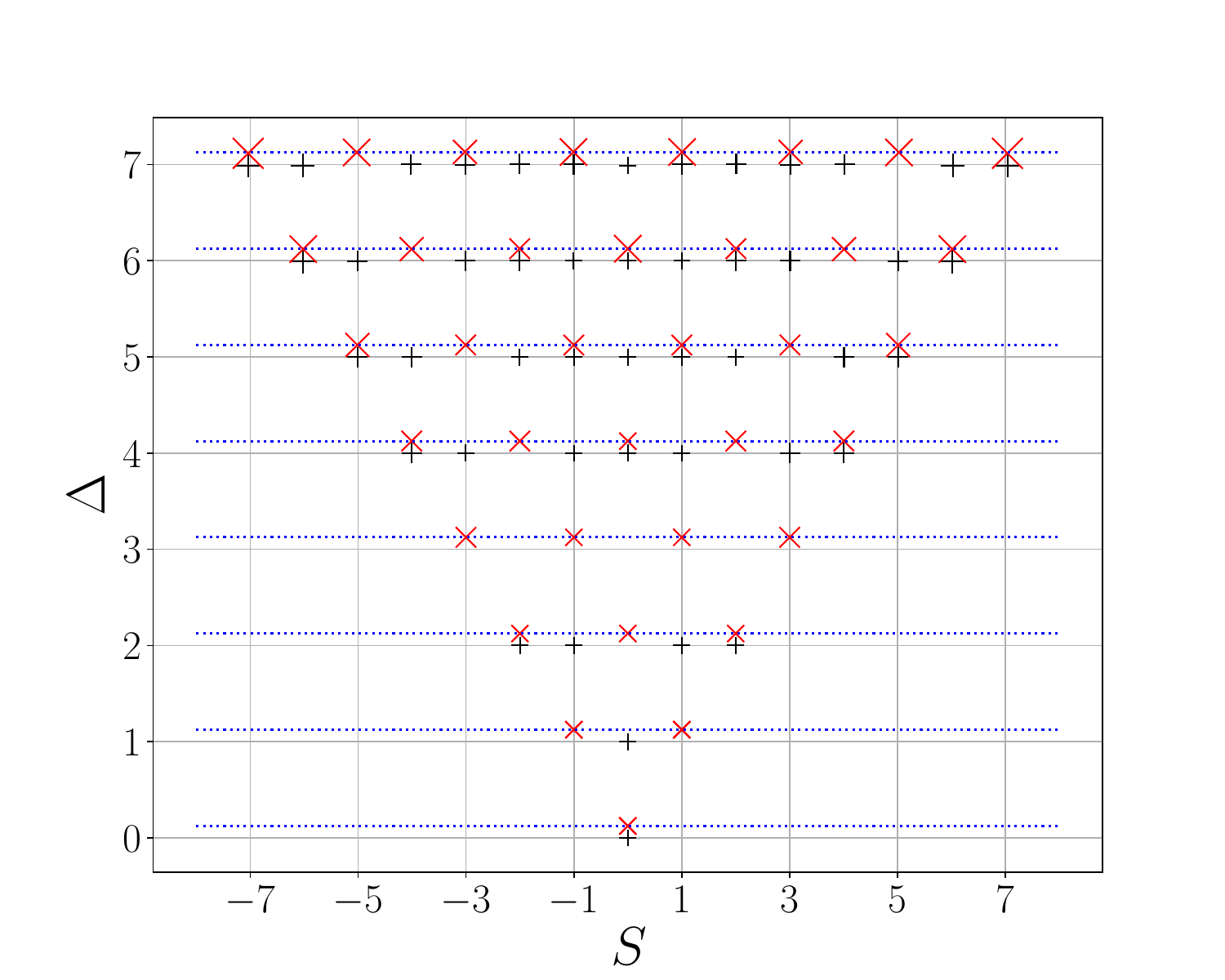}
    \caption{Conformal tower of the Ising model. The plot is based on HOTRG results with $D=100$ and $n=6$.}
    \label{fig:tower_Ising}
  \end{center}
\end{figure}

\begin{figure}[t]
  \begin{center}
    \includegraphics[width=0.95\columnwidth]{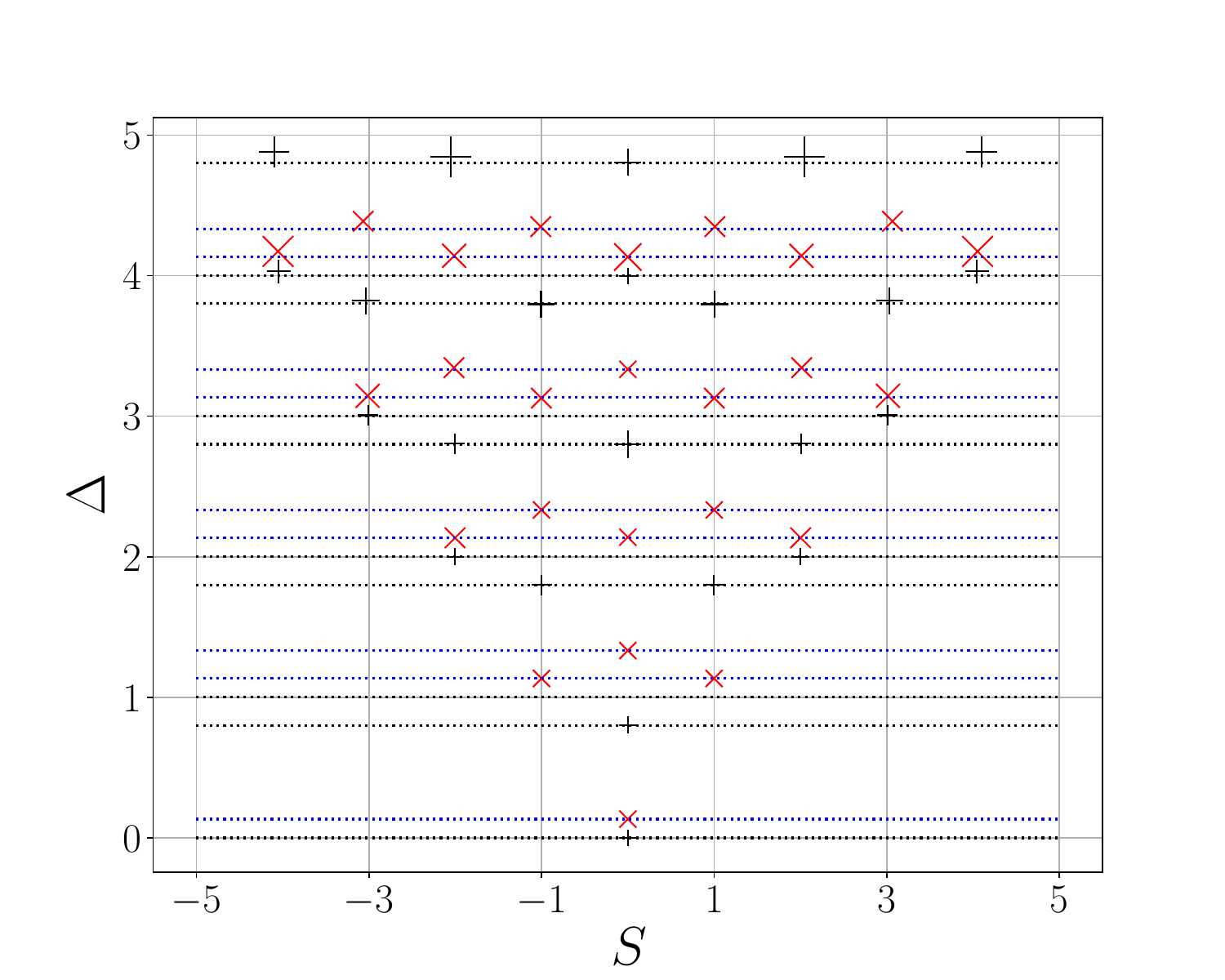}
    \caption{Conformal tower of the 3-state clock model. The plot is based on HOTRG results with $D=100$ and $n=6$.}
    \label{fig:tower_3clock}
  \end{center}
\end{figure}

We now present the main numerical results for the Ising and 3-state clock models.
Figures~\ref{fig:tower_Ising} and \ref{fig:tower_3clock} show the conformal towers extracted from HOTRG calculations with bond dimension $D=100$ and stacking number $n=6$.
The scaling dimensions and conformal spins are obtained from the finite-size analysis introduced above and refined in the model-by-model discussions below.
Already at this level, the comparison with the exact CFT spectra shows that the finite-size tensor-network flow captures a substantial portion of the low-lying conformal structure.
For the Ising model, we find agreement up to scaling dimension about $7$ in the $Q=0$ sector and $7\frac{1}{8}$ in the $Q=1$ sector.
For the 3-state clock model, the agreement extends up to about $4\frac{4}{5}$ in the $Q=0$ sector and $4\frac{1}{3}$ in the $Q=\pm 1$ sectors.

To quantify this agreement, we compare each identified level with the set of allowed CFT operators and check three quantities: 
the conformal spin $S_\phi$, the scaling dimension $\Delta_\phi$, and the degeneracy $d(\phi)$.
This comparison is used only as an a posteriori validation step; the numerical procedure itself does not assume detailed prior knowledge of the operator content.
We therefore begin by summarizing the exact CFT data needed for the validation and for the relative-error estimates quoted below.

CFT operators consist of primaries and their descendants.
The primaries are labeled by $(h,\bar{h})$, where $h$ and $\bar{h}$ are the holomorphic and antiholomorphic conformal weights, 
while descendants are specified by two nonnegative integers $(m, \bar{m})$.
For an operator $\phi$ labeled by $(h,\bar{h})+(m,\bar{m})$, the scaling dimension and conformal spin are
$\Delta_\phi=h+m+\bar{h}+\bar{m}$, $S_\phi=h+m-\bar{h}-\bar{m}$,
and the corresponding degeneracy is $d(\phi)=d(h,m)\times d(\bar{h},\bar{m})$.

For benchmarking, we use the exact primary content and descendant degeneracies of the Ising CFT in the two symmetry sectors relevant here, $Q=0$ and $Q=1$.
The corresponding primary data and degeneracy tables are summarized in Tables~\ref{tab:Ising_primaries} and \ref{deg_Ising}.
The full lists of allowed operators within the numerical window, together with the extracted crossover scales $L_x^*$ and relative errors $\mathrm{RE}_\Delta$, are deferred to Appendix~\ref{sec:conformal_data_Ising}.

We use the 3-state clock CFT data in the same way: Tables~\ref{tab:3clock_primaries} and \ref{deg_3clock} summarize the exact primaries and descendant degeneracies needed for validation.
The full operator lists within the numerical window, along with the corresponding values of $L_x^*$ and $\mathrm{RE}_\Delta$, are given in Appendix~\ref{sec:conformal_data_3clock}.
In the main text, we only refer to the broad sector structure, namely $Q=0$ and $Q=\pm1$, and return to operator-level details only when they are needed to interpret a specific numerical feature.

\begin{table}[t]
\caption{\label{tab:Ising_primaries} Primary operators of Ising model CFT}
\begin{ruledtabular}
\begin{tabular}{ccccc}
Operator & $\left(h,\bar{h}\right)$ & $\Delta=h+\bar{h}$ & $S=h-\bar{h}$ & $Z_2$ charge \\  \hline
$\mathbf{I}$ & $(0,0)$ & $0$ & $0$ & 0\\ 
$\sigma$ & $\left(\frac{1}{16},\frac{1}{16}\right)$ & $\frac{1}{8}$ & $0$ & 1 \\ 
$\epsilon$ & $\left(\frac{1}{2},\frac{1}{2}\right)$ & $1$ & $0$ & 0 \\ 
\end{tabular}
\end{ruledtabular}
\end{table}

\begin{table}[b]
\caption{\label{deg_Ising} $d(h,m)$ of Ising model CFT}
\begin{ruledtabular}
\begin{tabular}{lcccccccc}
$m$                      & 0 & 1 & 2 & 3 & 4 & 5 & 6 & 7\\  \hline
$h=0$                   & 1 & 0 & 1 & 1 & 2 & 2 &3 &3 \\ \hline 
$h=\frac{1}{16} $  & 1 & 1 & 1 & 2 & 2 & 3 &4&5 \\ \hline 
$h=\frac{1}{2}   $  & 1 & 1 & 1 & 1 & 2 & 2 &3 &4\\  
\end{tabular}
\end{ruledtabular}
\end{table}

\begin{table}[t]
\caption{\label{tab:3clock_primaries} Primary operators of 3-state clock model CFT}
\begin{ruledtabular}
\begin{tabular}{ccccc}
Operator & $\left(h,\bar{h}\right)$ & $\Delta=h+\bar{h}$ & $S=h-\bar{h}$ & $Z_3$ charge \\  \hline
$\mathbf{I}$ & $(0,0)$ & $0$ & $0$ & $0$ \\ 
$\epsilon$ & $\left(\frac{2}{5},\frac{2}{5}\right)$ & $\frac{4}{5}$ & $0$ & $0$ \\ 
$\Phi_{X\bar{\epsilon}}$ & $\left(\frac{2}{5},\frac{7}{5}\right)$ & $\frac{9}{5}$ & $-1$ & $0$ \\ 
$\Phi_{\epsilon\bar{X}}$ & $\left(\frac{7}{5},\frac{2}{5}\right)$ & $\frac{9}{5}$ & $+1$ & $0$ \\ 
$X$ & $\left(\frac{7}{5},\frac{7}{5}\right)$ & $\frac{14}{5}$ & $0$ & $0$ \\ 
$W$ & $(3,0)$ & $3$ & $+3$ & $0$ \\ 
$\bar{W}$ & $(0,3)$ & $3$ & $-3$ & $0$ \\ 
$Y$ & $(3,3)$ & $6$ & $0$ & $0$ \\ 
$\sigma$ & $\left(\frac{1}{15},\frac{1}{15}\right)$ & $\frac{2}{15}$ & $0$ & $+1$ \\ 
$\sigma^{\dagger}$ & $\left(\frac{1}{15},\frac{1}{15}\right)$ & $\frac{2}{15}$ & $0$ & $-1$ \\ 
$Z$ & $\left(\frac{2}{3},\frac{2}{3}\right)$ & $\frac{4}{3}$ & $0$ & $+1$ \\ 
$Z^{\dagger}$ & $\left(\frac{2}{3},\frac{2}{3}\right)$ & $\frac{4}{3}$ & $0$ & $-1$ \\ 
\end{tabular}
\end{ruledtabular}
\end{table}

\begin{table}[b]
\caption{\label{deg_3clock} $d(h,m)$ of 3-state clock model CFT} 
\begin{ruledtabular}
\begin{tabular}{ccccccc}
Level                    & 0 & 1 & 2 & 3 & 4 & 5 \\  \hline
$h=0$                  & 1 & 0 & 1 & 1 & 2 & 2 \\ \hline 
$h=\frac{1}{15}$  & 1 & 1 & 2 & 3 & 5 & 7 \\ \hline 
$h=\frac{2}{5}  $  & 1 & 1 & 1 & 2 & 3 & 4 \\  \hline 
$h=\frac{7}{5}  $  & 1 & 1 & 2 & 2 & 4 & 5 \\  \hline 
$h=\frac{2}{3}  $  & 1 & 1 & 2 & 2 & 4 & 5 \\ 
\end{tabular}
\end{ruledtabular}
\end{table}

Recall that the goal is to extract conformal data from tensor-network calculations without assuming detailed prior knowledge of the underlying CFT.
We use only general structural information: states can be organized by their $\mathcal{Z}_n$ charge $Q$, 
allowed scaling dimensions are separated by finite gaps within each charge sector, and conformal spins are integers, up to the expected spin-flip symmetry.
Accordingly, the finite-size levels $X_k^Q(L_x)$ should first organize into nearly degenerate groups associated with a common scaling dimension $\Delta$, 
and these groups can then be further resolved into spin sectors using the corresponding values of $S_k(L_x)$.
The total multiplicity of a group is fixed by the allowed operator content, while the multiplicity of each spin-resolved subgroup is fixed by the subset of operators with the same $(Q,\Delta,S)$.

\begin{figure}[t]
  \includegraphics[width=0.95\columnwidth]{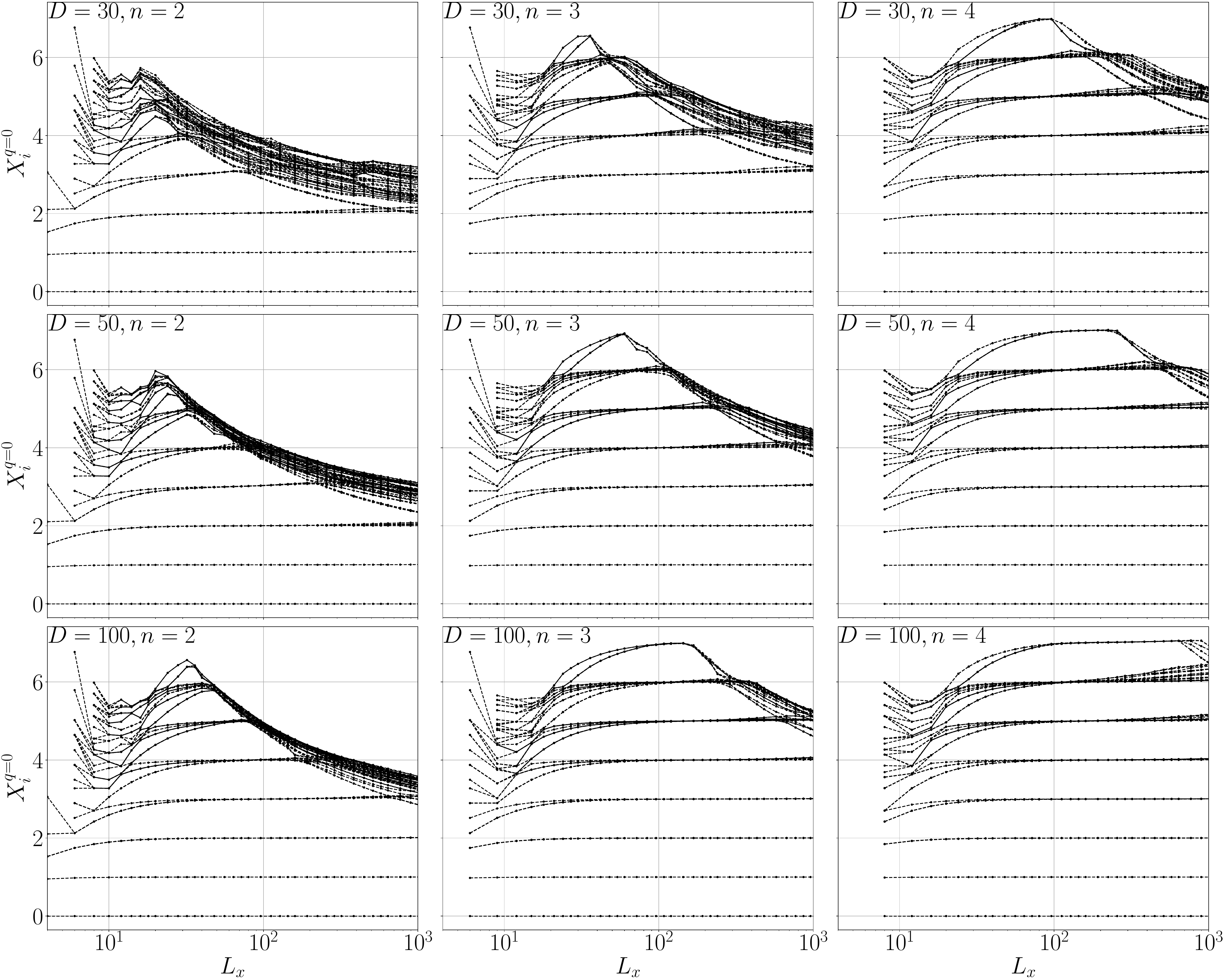}
  \caption{HOTRG results for the Ising model for $X^{Q=0}_i$ as a function of $L_x$, with $i=0\cdots 60$.
  From top to bottom, the rows correspond to bond dimensions $D=30, 50, 100$, respectively, 
  while from left to right, the columns correspond to $n=2,3,4$, respectively.}
\label{fig:HOTRG_Xi_q0}
\end{figure}

\begin{figure}[t]
\begin{center}
 \includegraphics[width=0.95\columnwidth]{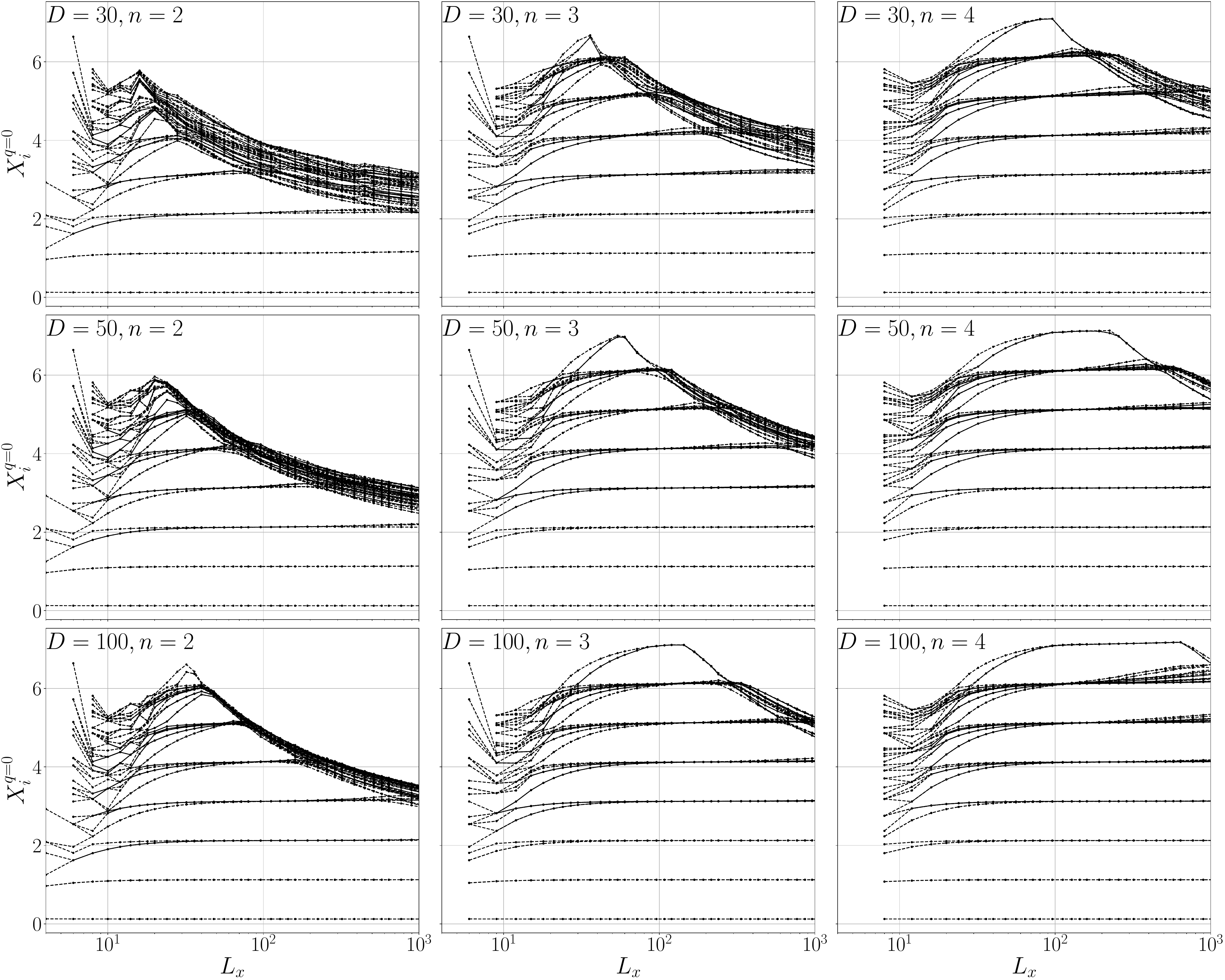}
 \caption{HOTRG results of Ising model for $X^{Q=1}_i$ as a function of $L_x$, with $i=0\cdots 60$.
  From top to bottom, the rows correspond to bond dimensions $D=30, 50, 100$, respectively, 
  while from left to right, the columns correspond to $n=2,3,4$, respectively.}
\label{fig:HOTRG_Xi_q1}
\end{center}
\end{figure}

In tensor-network calculations, this ideal structure is visible only up to a method- and bond-dimension-dependent length scale.
Below that scale, the numerically extracted $X_k^Q(L_x)$ and $S_k(L_x)$ remain close to their CFT values; beyond it, truncation effects drive them away from the critical finite-size flow.
We illustrate this behavior first for the Ising model.
Figures~\ref{fig:HOTRG_Xi_q0} and \ref{fig:HOTRG_Xi_q1} show the lowest sixty finite-size scaling dimensions $X_k^Q(L_x)$ in the $Q=0$ and $Q=1$ sectors, 
obtained with HOTRG for bond dimensions $D=30,50,100$ and stacking numbers $n=2,3,4$.

Focus first on the bottom-right panel, corresponding to $D=100$ and $n=4$.
As $L_x$ increases, the levels $X_k^Q(L_x)$ organize into groups separated by clear gaps, as expected from the underlying conformal tower.
Higher-scaling-dimension groups emerge only at larger system sizes, and they begin to broaden again once $L_x$ becomes too large.
This broadening signals the growing effect of the finite bond dimension.
From the renormalization-group perspective, a finite-$D$ truncation acts as a relevant perturbation \cite{Ueda.2023, Huang.2023, Maiti.2025}, 
driving the flow away from the critical fixed point and causing $X_k^Q(L_x)$ to deviate from their exact values.

Upon decreasing either $D$ or $n$, the same overall grouping pattern remains visible, but the highest-scaling-dimension groups collapse at smaller system sizes.
As $L_x$ increases further, this collapse progressively propagates to groups with lower scaling dimensions, making the grouping ambiguous beyond that point.
Such behavior has been reported previously \cite{Iino.2019ud, Iino.2020}, although its microscopic origin remains unclear.
Our results show that the onset of the collapse moves to smaller $L_x$ as either $D$ or $n$ is reduced, with the stacking number $n$ producing the stronger effect.

\begin{figure*}[t]
\begin{center}
 \includegraphics[width=\textwidth]{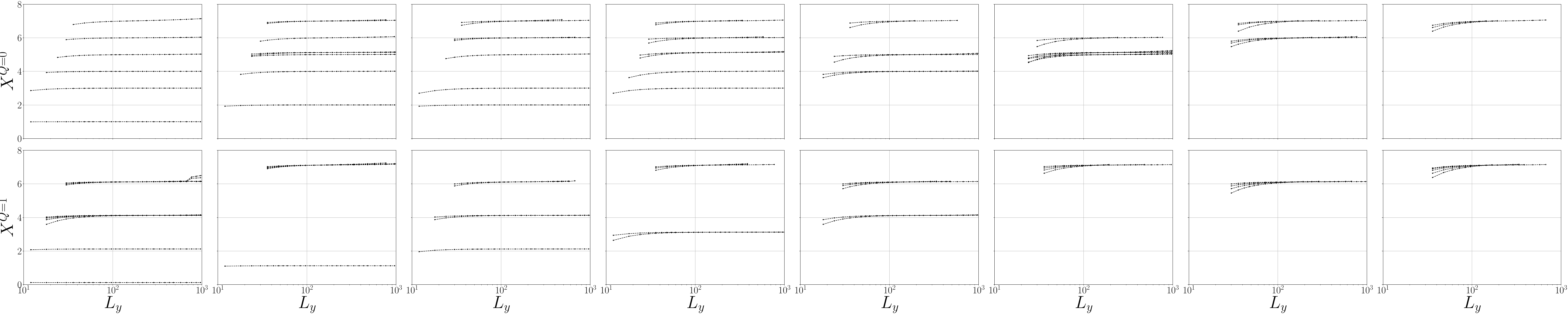}
\caption{HOTRG results for the Ising model for self-consistent spin-resolved subgroups $X^{Q,S}_k$ as a function of $L_x$ with $D=100, n=6$. 
The first (second) row corresponds to $Q=0$ ($Q=1$), while columns from left to right correspond to $S=0$ to $S=7$.}
\label{fig:Ising_q0_q1}
\end{center}
\end{figure*}

We also find that $S_k^Q(L_x)$ remains very close to an integer up to a nonuniversal system size and then begins to drift away.
This provides a practical self-consistency criterion: once the deviation of $S_k^Q(L_x)$ from the nearest integer exceeds a chosen threshold, we discard the data beyond that size.
Within the remaining self-consistent window, we use the conformal spin to split each group into spin-resolved subgroups.
Figure~\ref{fig:Ising_q0_q1} shows the resulting subgroups $X_k^{Q,S}(L_x)$ for $Q=0,1$ and $S=0,1,\cdots,7$, 
obtained from HOTRG with $D=100$ and $n=6$; only nonnegative $S$ is shown because the $\pm S$ sectors are essentially identical.
For the Ising model, we can identify such subgroups up to $\Delta \approx 7$ in the $Q=0$ sector and $\Delta \approx 7.125$ in the $Q=1$ sector.
This identification step is central to the framework: whenever a spin-resolved subgroup can be isolated, its conformal spin, scaling dimension, and degeneracy can be estimated accurately.

For a self-consistent spin-resolved subgroup, we estimate the scaling dimension from its average value $\bar{X}_S^Q(L_x)$.
Following Refs.~\cite{Huang.2023, Maiti.2025}, we define the crossover scale $L_x^*$ by the condition that $\left| \frac{d}{dl} \, \bar{X}_S^Q(L_x) \right|$ is minimal, with $l=\ln L_x$, and take $\bar{X}_S^Q(L_x^*)$ as the optimal estimate of $\Delta$.

\subsection{Ising model \label{sec:Ising}}

Focusing first on the Ising model,
we find that the finite-size tensor-network flow cleanly resolves a broad hierarchy of spin-resolved subgroups and provides highly accurate conformal data over a wide scaling window.
To make the logic of the analysis transparent, we begin with simple low-lying groups and then move to progressively more challenging cases in which degeneracies, crossover identification, and shrinking self-consistent windows all become more important.

\subsubsection{$X^{Q=1}_{0} \approx 0.125$}

\begin{figure}[b]
\begin{center}
 \includegraphics[width=0.95\columnwidth]{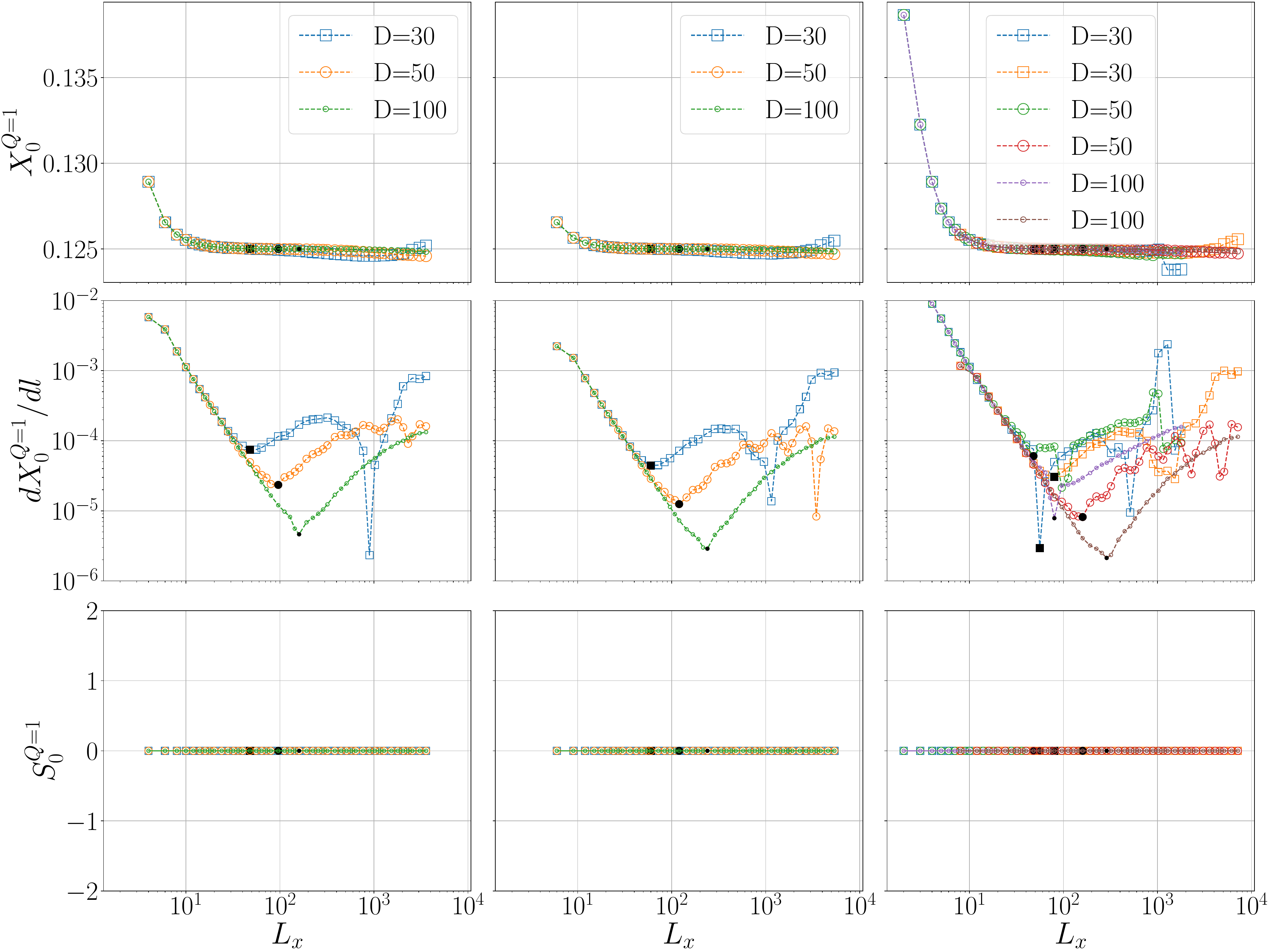}
  \caption{First row: $X^{Q=1}_{0}$ as a function of $L_x$.
                Second row: $\left| \frac{d}{dl}X^{Q=1}_0 \right|$ as a function of $L_x$.
                Third row: $S^{Q=1}_0$ as a function of $L_x$.
                The columns correspond to different values of $n$, with $n=2,3,4$ from left to right.
                Solid symbols correspond to results at $L_x^*(D,n)$.}
\label{fig:Ising_sigma}
\end{center}
\end{figure}

We begin with the simplest case: the lowest group in the $Q=1$ sector, which consists only of $X^{Q=1}_0$.
Its flow already illustrates the basic structure of the analysis: $X^{Q=1}_0(L_x)$ approaches $0.125$ while $S^{Q=1}_0(L_x)$ remains consistent with zero.
In Fig.~\ref{fig:Ising_sigma}, the first and third rows show $X^{Q=1}_0$ and $S^{Q=1}_0$ on a semi-log scale for $n=2,3,4$ and $D=30,50,100$.
Across all parameter choices, the conformal spin remains within the self-consistency threshold, so no data need to be discarded.
At the same time, the scaling-dimension data collapse onto a common curve up to a $D$- and $n$-dependent crossover scale, beyond which the flow becomes nonuniversal.

Following Ref.~\cite{Huang.2023, Maiti.2025} we identify the crossover length scale $L_x^*$ 
by plotting $\left| \frac{d} {dl}  X^{Q=1}_{0}\right|, l=\ln L_x$ in the second row.
We find that it follows a universal power-law decay up to some $L_x^*$, while for $L_x>L_x^*$ it starts to increase in a non-universal way.
Consequently, we identify $L_x^*$ as the system size at which it reaches its minimum.
In general $L_x^*$ depends on both $D$ and $n$. 
Our results clearly show that $L_x^*$ increases when $D$ increases for a fixed $n$, or when $n$ increases for a fixed $D$.
Moreover, we find that increasing $n$ has a stronger effect in this parameter range.
Finally, we use the results at $L_x^*(D,n)$ as the best estimate of the scaling dimension for a given set of $D$ and $n$.

It is important to note that the above finite-size scaling analysis does not rely on a priori knowledge of the underlying CFT.
By analyzing this group, we obtain one nondegenerate point in the conformal tower
\begin{equation}
  \left( S_\phi=0, \Delta_\phi \approx X^{Q=1}_0 (L_x^*) \right),
\end{equation}
where $X^{Q=1}_0 (L_x^*) \approx 0.125$.

Now we compare the result to the exact CFT to evaluate the relative error.
We identify this level with the Ising spin operator $\phi=\sigma\equiv (\frac{1}{16},\frac{1}{16})$, for which $\Delta_\sigma=\frac{1}{8}$ and $S_\sigma=0$.
Within our framework, the conformal spin is quantized to an integer, so once the numerical value is consistent with the exact CFT, there is no residual spin error to report.
We therefore quantify the remaining discrepancy through the relative error of the scaling dimension,
\begin{equation}
  \text{RE}_\Delta \equiv \frac{ \left| X^{Q=1}_0 (L_x^*)-\Delta_\phi \right| }{\Delta_\phi}.
\end{equation}
Even for $D=30$ and $n=2$, the relative error is already as small as $1.1\times 10^{-4}$, and it decreases systematically as either $D$ or $n$ is increased.
At $D=100$ and $n=6$, the error is further reduced to $3.7 \times 10^{-6}$.
This example shows that low-lying scaling dimensions can be extracted with high accuracy at moderate bond dimension and with extremely high accuracy once the computational budget is pushed further.

\subsubsection{$X^{Q=1}_{1,2} \approx 1.125$}

\begin{figure}[t]
\begin{center}
 \includegraphics[width=0.95\columnwidth]{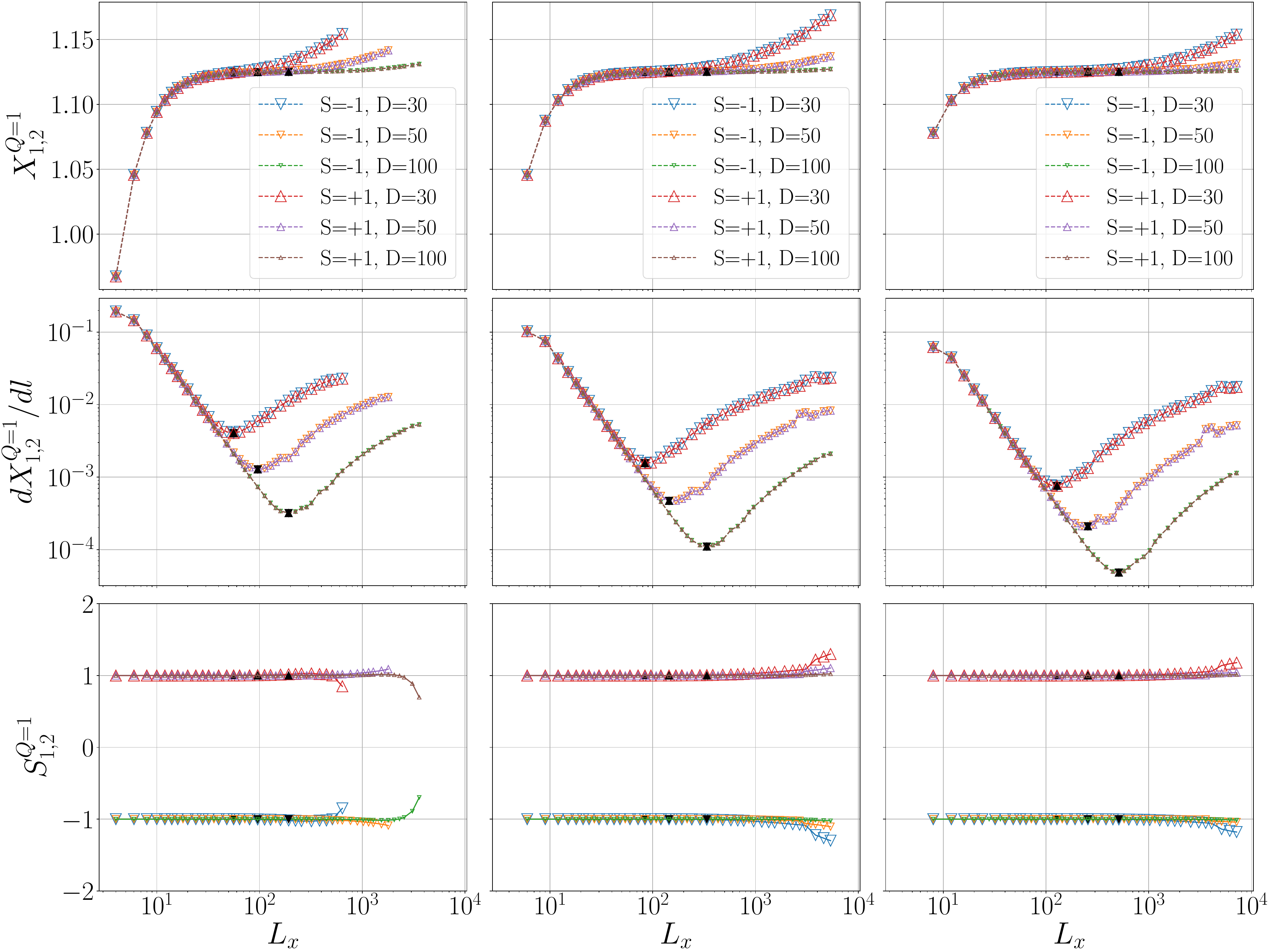}
  \caption{First row: $X^{Q=1}_{1,2}$ as a function of $L_x$.
                Second row: $\left| \frac{d}{dl}X^{Q=1}_{1,2} \right|$ as a function of $L_x$.
                Third row: $S^{Q=1}_{1,2}$ as a function of $L_x$.
                The columns correspond to different values of $n$, with $n=2,3,4$ from left to right.
                Solid symbols correspond to results at $L_x^*(D,n)$.}
\label{fig:Ising_q1_X2_3}
\end{center}
\end{figure}

We next consider the second group in the $Q=1$ sector, whose scaling dimension is close to $1.125$ and which contains two nearly degenerate states.
Here the key new feature is that conformal spin resolves the near degeneracy cleanly, splitting the group into two spin-resolved subgroups with $S=\pm 1$.
This immediately yields an unambiguous assignment of $X^{Q=1}_1$ and $X^{Q=1}_2$ to the $S=+1$ and $S=-1$ branches, respectively.
The first and third rows of Fig.~\ref{fig:Ising_q1_X2_3} show the corresponding flows of $X^{Q=1}_{1,2}$ and $S^{Q=1}_{1,2}$, with the expected spin-flip symmetry clearly visible.

Following the standard procedure, 
in the second row we plot $\left| \frac{d } {dl} X^{Q=1}_{1,2} \right|$ to identify $L_x^*$.
Again, we find universal behavior when $L<L_x^*$ and $L_x^*$ increases with increasing $D$ or $n$.
We also observe that $S^{Q=1}_{1,2}$ start to deviate from $\pm 1$ at large system size.
But the deviation is still within the threshold and no data is dropped.
Moreover, in all cases the  onset of the deviation is larger than corresponding $L_x^*$.
Consequently, the deviation does not interfere with the estimation of the scaling dimension.

The analysis for this group results in two non-degenerate points in the conformal tower:
\begin{eqnarray}
  && \left( S_{\phi_1}=+1, \Delta_{\phi_1} \approx X^{Q=1}_1 (L_x^*) \right), \\
  && \left( S_{\phi_2}=-1, \Delta_{\phi_2} \approx X^{Q=1}_2 (L_x^*)  \right),
\end{eqnarray}
where $X^{Q=1}_{1,2} \approx 1.125$.
Comparison with the Ising CFT identifies these two levels as
$\phi_1 = \left(\frac{1}{16},\frac{1}{16}\right)+(1,0)$ and
$\phi_2 = \left(\frac{1}{16},\frac{1}{16}\right)+(0,1)$.
As in the previous example, the accuracy improves systematically with increasing $D$ or $n$, reaching relative errors of about $2.2 \times 10^{-6}$ at $D=100$ and $n=6$.

\subsubsection{$X^{Q=1}_{6, \dots ,11} \approx 3.125$}

We now turn to a case in which spin resolution will not remove all degeneracies.
The group near $\Delta\approx 3.125$ in the $Q=1$ sector contains six nearly degenerate states, $X^{Q=1}_{6}$ through $X^{Q=1}_{11}$.
Using conformal spin, we resolve this manifold into four subgroups with $S=\pm 3$ and $\pm 1$.
The $S=\pm 1$ branches are nondegenerate, whereas the $S=\pm 3$ branches remain doubly degenerate.
We therefore assign $X^{Q=1}_{6,7}$, $X^{Q=1}_{8}$, $X^{Q=1}_{9}$, and $X^{Q=1}_{10,11}$ to the $S=+3$, $+1$, $-1$, and $-3$ subgroups, respectively, and work with averaged quantities $\overline{X}^{Q=1}_{\pm 3}$ and $\overline{S}^{Q=1}_{\pm 1}$ to keep the notation uniform.

\begin{figure}[t]
\begin{center}
 \includegraphics[width=0.95\columnwidth]{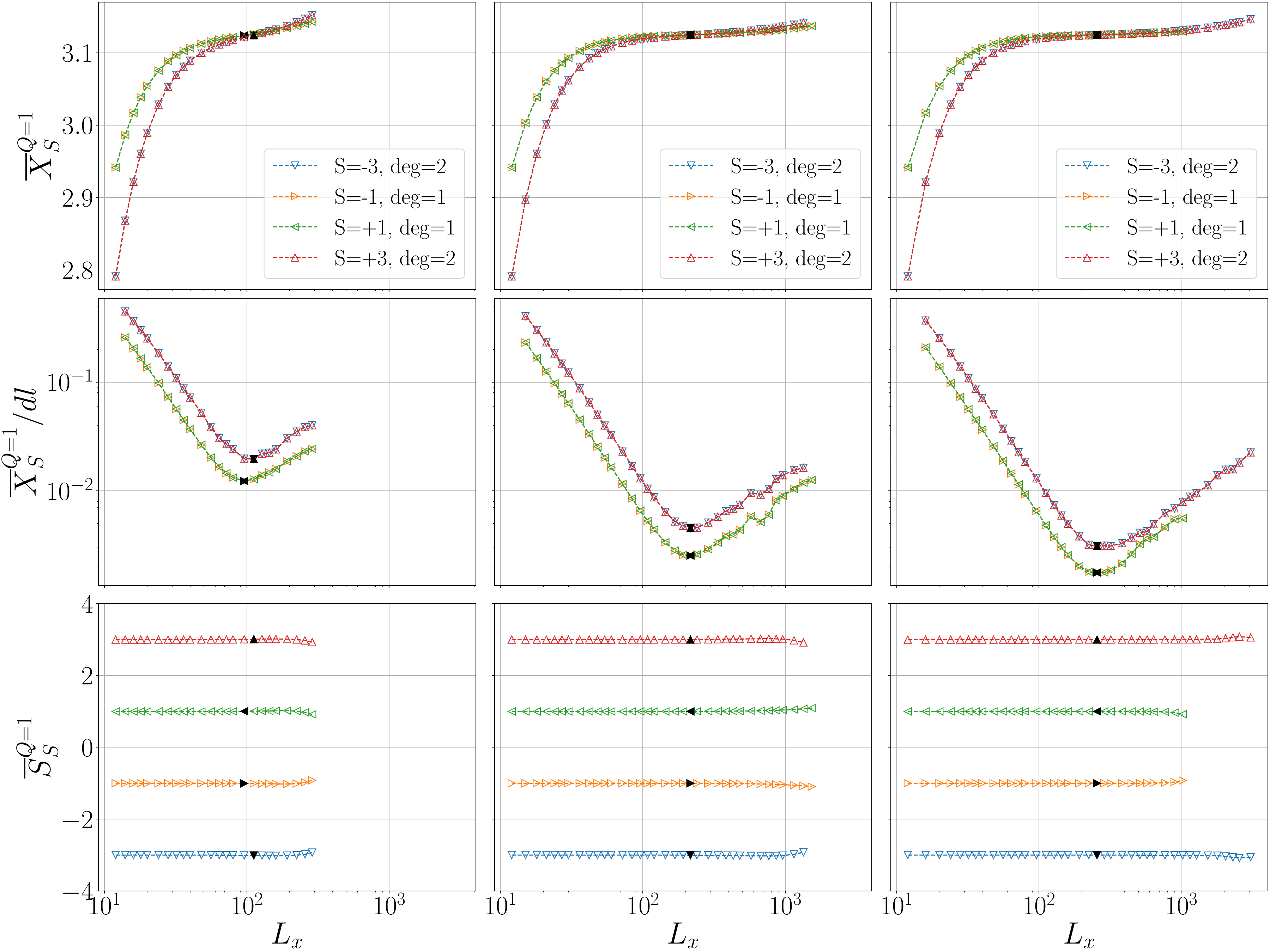}
  \caption{First row: $\overline{X}^{Q=1}_{S=\pm 3, \pm 1}$ btained from $X^{Q=1}_{6, \dots, 11}$  as a function of $L_x$.
                Second row: $\left| \frac{d}{dl} \overline{X}^{Q=1}_{S=\pm 3, \pm 1} \right|$ as a function of $L_x$.
                Third row: $S^{Q=1}_{S=\pm 3, \pm 1}$ as a function of $L_x$.
                The columns correspond to different values of $n$, with $n=2,3,4$ from left to right.
                Solid symbols correspond to results at $L_x^*(D,n)$.}
\label{fig:Ising_3.125}
\end{center}
\end{figure}

In the first row of Fig.~\ref{fig:Ising_3.125} we plot $\overline{X}^{Q=1}_{\pm 3, \pm 1}$ and label their degeneracy in the legend.
Here we only plot $D=100$ results for clarity, but the data shows that 
increasing $D$ or $n$ can systemically improve the accuracy.
As expected the spin-flip symmetry is observed.
We note that level-crossing between $\overline{X}^{Q=1}_{\pm 3}$ and $\overline{X}^{Q=1}_{\pm 1}$ is observed. 
This indicates that it is important to use conformal spin to lift the degeneracy.
In the third row of Fig.~\ref{fig:Ising_3.125} we plot $\overline{S}^{Q=1}_{\pm S}$.
Here we observe that with $n=2$ the conformal spin starts to deviate from the integer value at much smaller system size,
indicating larger $n$ is needed to ensure a wide self-consistent regime.

In second row we plot  $\left| \frac{d}{dl} \overline{X}^{Q=1}_{\pm S} \right|$ and use it to identify the associated crossover length $L_x^*$.
From the analysis we obtain 4 points in the conformal tower
\begin{eqnarray}
  && \left( S_{\phi_1}=+3, \Delta_{\phi_1} \approx \overline{X}^{Q=1}_{S=+3} (L_x^*) \right), \text{deg}=2, \\
  && \left( S_{\phi_2}=+1, \Delta_{\phi_2} \approx \overline{X}^{Q=1}_{S=+1} (L_x^*) \right), \text{deg}=1, \\
  && \left( S_{\phi_3}=-1, \Delta_{\phi_3} \approx \overline{X}^{Q=1}_{S=-1} (L_x^*)  \right), \text{deg}=1, \\
  && \left( S_{\phi_4}=-3, \Delta_{\phi_4} \approx \overline{X}^{Q=1}_{S=-3} (L_x^*) \right), \text{deg}=2,
\end{eqnarray}
where $\overline{X}^{Q=1}_{\pm 3, \pm 1}(L_x^*) \approx 3.125$.
Comparison with the Ising CFT leads to the identification
$\phi_1 = \left(\frac{1}{16},\frac{1}{16}\right)+(3,0)$,
$\phi_2 = \left(\frac{1}{16},\frac{1}{16}\right)+(2,1)$,
$\phi_3 = \left(\frac{1}{16},\frac{1}{16}\right)+(1,2)$, and
$\phi_4 = \left(\frac{1}{16},\frac{1}{16}\right)+(0,3)$.
The degeneracy pattern is also reproduced correctly, with $d(\phi_2)=d(\phi_3)=1$ and $d(\phi_1)=d(\phi_4)=2$.
For $D=100$ and $n=6$, the relative errors are $3.9 \times 10^{-5}$ for $\Delta_{\phi_1,\phi_4}$ and $5.5 \times 10^{-5}$ for $\Delta_{\phi_2,\phi_3}$, showing that the method remains accurate even when residual degeneracies are present after spin resolution.

\subsubsection{$X^{Q=1}_{57, \dots, 88} \approx 7.125$}

\begin{figure}[t]
\begin{center}
 \includegraphics[width=0.95\columnwidth]{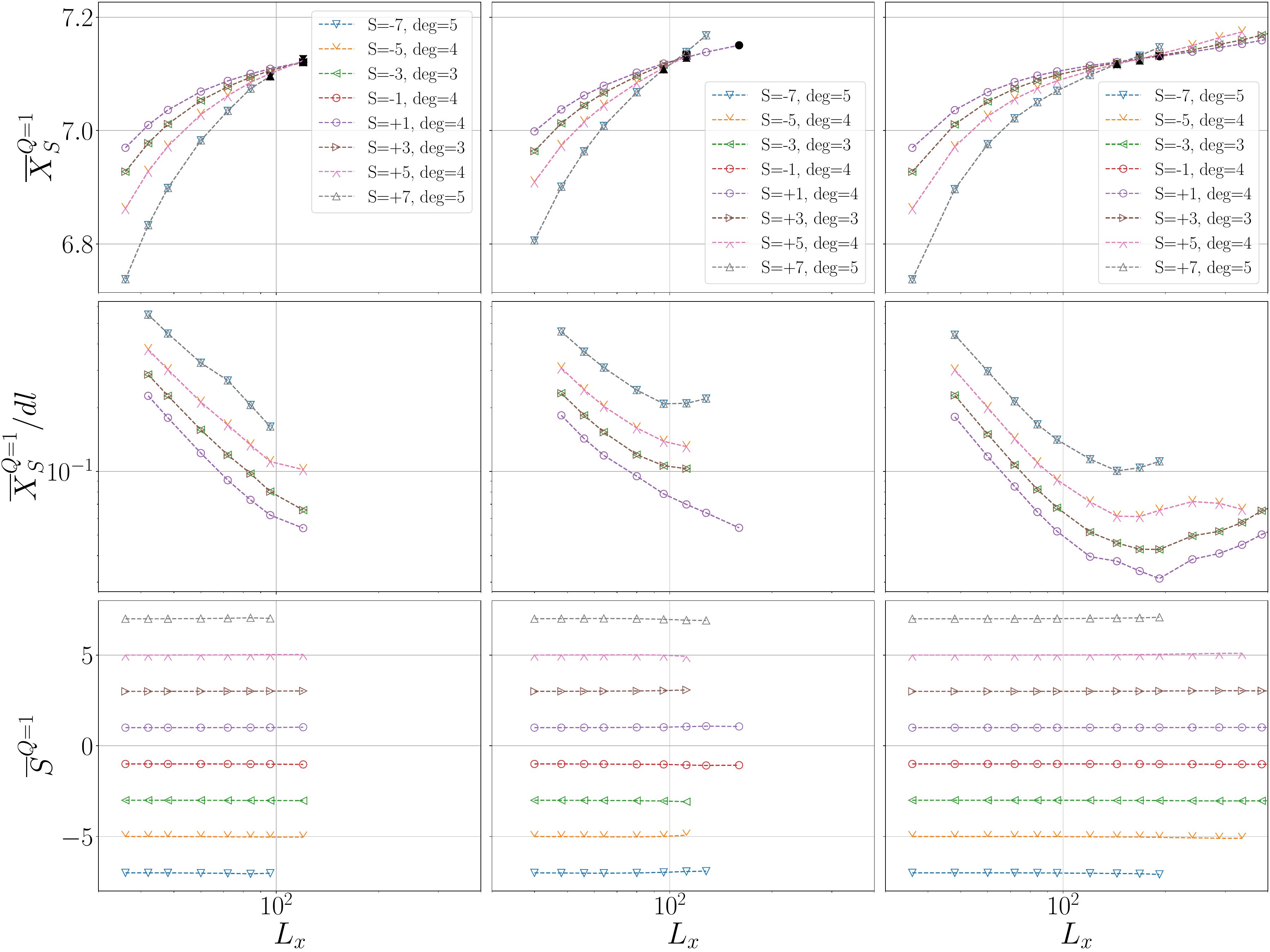}
  \caption{First row: $\overline{X}^{Q=1}_{S}$ obtained from $X^{Q=1}_{57, \dots, 88}$ as a function of $L_x$.
                Second row: $\left| \frac{d}{dl}X^{Q=1}_{S} \right|$ as a function of $L_x$.
                Third row: $S^{Q=1}_{S}$ as a function of $L_x$.
                The columns correspond to different values of $n$, with $n=3,4,6$ from left to right.
                Solid symbols correspond to results at $L_x^*(D,n)$.}
\label{fig:Ising_7.125}
\end{center}
\end{figure}

The final Ising example probes the edge of the numerically accessible window.
The highest-scaling-dimension group that we can still identify in the $Q=1$ sector lies near $\Delta\approx 7.125$ and contains 32 states, from $X^{Q=1}_{57}$ to $X^{Q=1}_{88}$.
Conformal spin partially resolves this large multiplet into eight subgroups with $S=\pm 7$, $\pm 5$, $\pm 3$, and $\pm 1$, whose degeneracies are 5, 4, 3, and 4, respectively.
The first row of Fig.~\ref{fig:Ising_7.125} shows the averaged quantities $\overline{X}^{Q=1}_{S}$ for these subgroups, using HOTRG data at $D=100$ and $n=3,4,6$.
It is important to observe the substantial shrinking of the self-consistent regime.
On the one hand, due to stronger intrinsic finite-size effects a larger system size is required before a high scaling dimension group can be identified.
On the other hand, high scaling dimension groups also suffer from stronger finite-bond-dimension-induced relevant perturbation.
These two effects result in the shrinking of the self-consistent regime.
Consequently, for $n=3$ and $4$ the crossover length $L_x^*$ is near the end of the self-consistent regime.
This suggests that the results will have lower accuracy.
By pushing $n$ to 6, we are able to identify a sufficiently large self-consistent regime within which $L_x^*$ can be identified reliably.
However, the value of $\left| \frac{d} {d l} X^{Q=1}_{S} (L_x^*)\right|$ is much larger than 
the corresponding values for the lower-scaling-dimension groups.
This suggests that the estimated scaling dimension will be less accurate than the estimates of lower scaling dimensions.

From the analysis of the $D=100, n=6$ results, we obtain 8 points in the conformal tower and the corresponding degeneracy:
\begin{eqnarray}
  && \left( S_{\phi_1}=+7, \Delta_{\phi_1} \approx \overline{X}^{Q=1}_{S=+7} (L_x^*) \right), \text{deg}=5, \\
  && \left( S_{\phi_2}=+5, \Delta_{\phi_2} \approx \overline{X}^{Q=1}_{S=+5} (L_x^*) \right), \text{deg}=4, \\
  && \left( S_{\phi_3}=+3, \Delta_{\phi_3} \approx \overline{X}^{Q=1}_{S=+3} (L_x^*) \right), \text{deg}=3, \\
  && \left( S_{\phi_4}=+1, \Delta_{\phi_4} \approx \overline{X}^{Q=1}_{S=+1} (L_x^*) \right), \text{deg}=4. \\
  && \left( S_{\phi_5}=-1, \Delta_{\phi_5} \approx \overline{X}^{Q=1}_{S=-1} (L_x^*) \right), \text{deg}=4. \\
  && \left( S_{\phi_6}=-3, \Delta_{\phi_6} \approx \overline{X}^{Q=1}_{S=-3} (L_x^*) \right), \text{deg}=3, \\
  && \left( S_{\phi_7}=-5, \Delta_{\phi_7} \approx \overline{X}^{Q=1}_{S=-5} (L_x^*) \right), \text{deg}=4, \\
  && \left( S_{\phi_8}=-7, \Delta_{\phi_8} \approx \overline{X}^{Q=1}_{S=-7} (L_x^*) \right), \text{deg}=5,
\end{eqnarray}
Comparison with the exact Ising spectrum then fixes the operator assignment for each subgroup, while the larger derivatives at $L_x^*$ confirm that the accuracy is visibly reduced relative to the lower-lying cases.

By matching the scaling dimension and conformal spin to the exact Ising CFT we can make the following identification: 
$\phi_1 = \left(\frac{1}{16},\frac{1}{16}\right)+(7,0)$,
$\phi_2 = \left(\frac{1}{16},\frac{1}{16}\right)+(6,1)$,
$\phi_3 = \left(\frac{1}{16},\frac{1}{16}\right)+(5,2)$,
$\phi_4 = \left(\frac{1}{16},\frac{1}{16}\right)+(4,3)$,
$\phi_5 = \left(\frac{1}{16},\frac{1}{16}\right)+(4,3)$,
$\phi_6 = \left(\frac{1}{16},\frac{1}{16}\right)+(5,2)$,
$\phi_7 = \left(\frac{1}{16},\frac{1}{16}\right)+(6,1)$, and
$\phi_8 = \left(\frac{1}{16},\frac{1}{16}\right)+(7,0)$.
Moreover, the exact degeneracy of these operators agrees with the degeneracy we numerically obtained.
Based on this identification we evaluate their relative errors.
With $D=100$ and $n=6$, the relative errors 
range from $1.1 \times 10^{-3}$ for $\phi_1, \phi_8$ to $1.8 \times 10^{-4}$ for $\phi_2, \phi_7$.
The value of the worst relative error is still quite small.

Now we are in a position to analyze all spin-resolved subgroups for the Ising model.
In Appendix~\ref{sec:conformal_data_Ising}, we plot the HOTRG results with $D=100$ and $n=6$ 
for all self-consistent spin-resolved subgroups in Fig.~\ref{fig:Ising_D100n6_q0}.
For each subgroup we use the above framework to estimate the conformal spin, scaling dimension, degeneracy, and crossover length scale.
Based on these results, we plot the conformal tower of the Ising model in Fig.~\ref{fig:tower_Ising}
and observe perfect agreement with the exact Ising CFT.
Finally, we list the crossover length scale $L_x^*$ and relative error $\text{RE}_\Delta(L_x^*)$ in Tables~\ref{tab:Ising_q0-1} and \ref{tab:Ising_q1-1} 
in Appendix~\ref{sec:conformal_data_Ising}.
As expected, higher scaling dimensions have larger relative errors.
With HOTRG at $D=100$ and $n=6$, the relative error is between $2.2 \times 10^{-6}$ and $1.1 \times 10^{-3}$.

In addition, in Tables~\ref{tab:Ising_q0-1} and \ref{tab:Ising_q1-1} we also list $\tilde{L}_x^*$, the largest available system size below $L_x^*$,
together with the relative error $\text{RE}_\Delta(\tilde{L}_x^*)$ of the CFT data extracted at $\tilde{L}_x^*$.
We find that $\text{RE}_\Delta(\tilde{L}_x^*)$ is consistently larger than $\text{RE}_\Delta(L_x^*)$ up to about level 6.
Beyond that level, $\text{RE}_\Delta(\tilde{L}_x^*)$ occasionally falls below $\text{RE}_\Delta(L_x^*)$ for some subgroups, but the difference is always small.
These results indicate that our identification of $L_x^*$ is nearly optimal.

\subsection{3-state clock model \label{sec:3-clock}}

\begin{figure}[t]
\begin{center}
 \includegraphics[width=0.95\columnwidth]{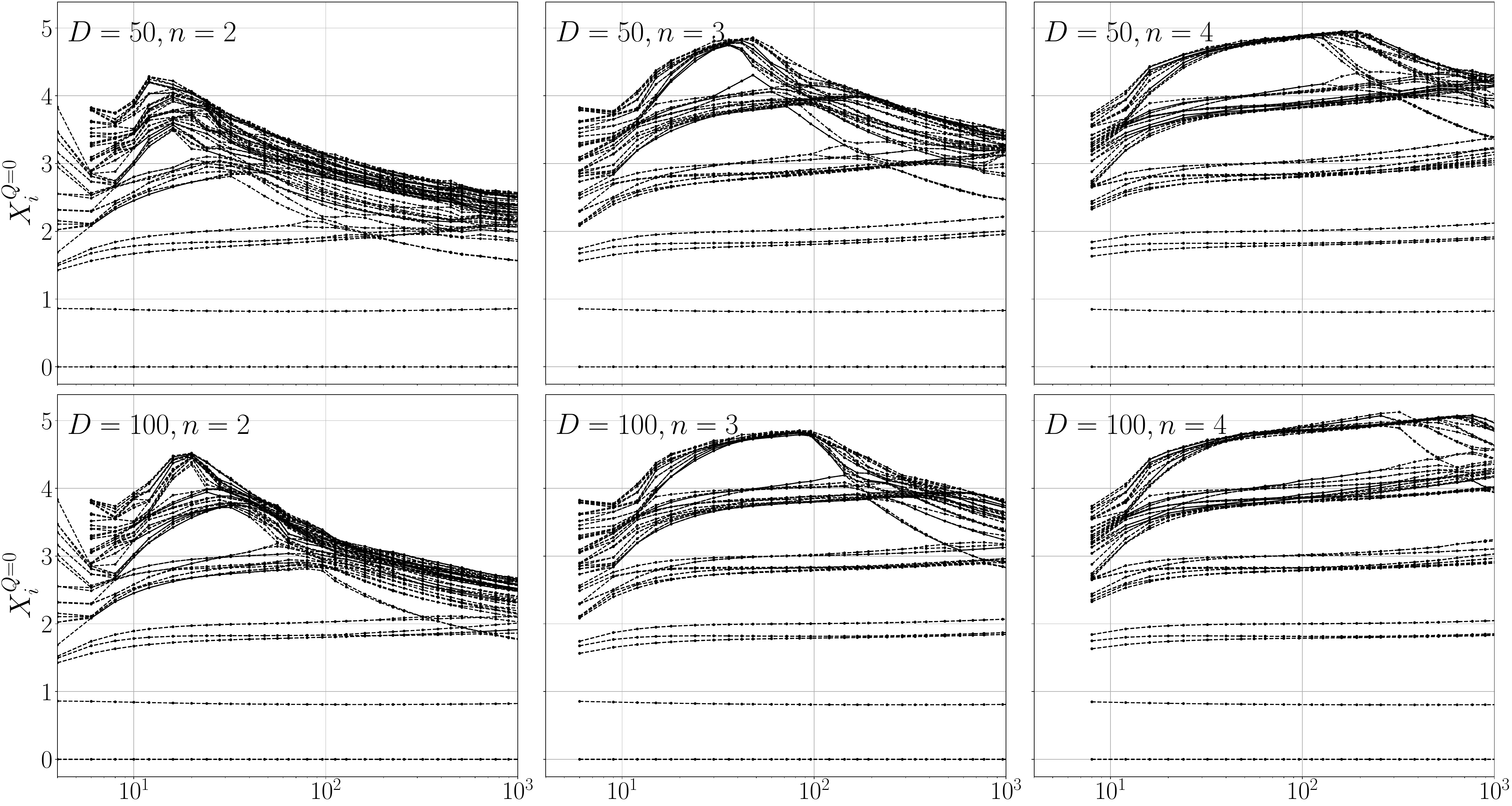}
 \includegraphics[width=0.95\columnwidth]{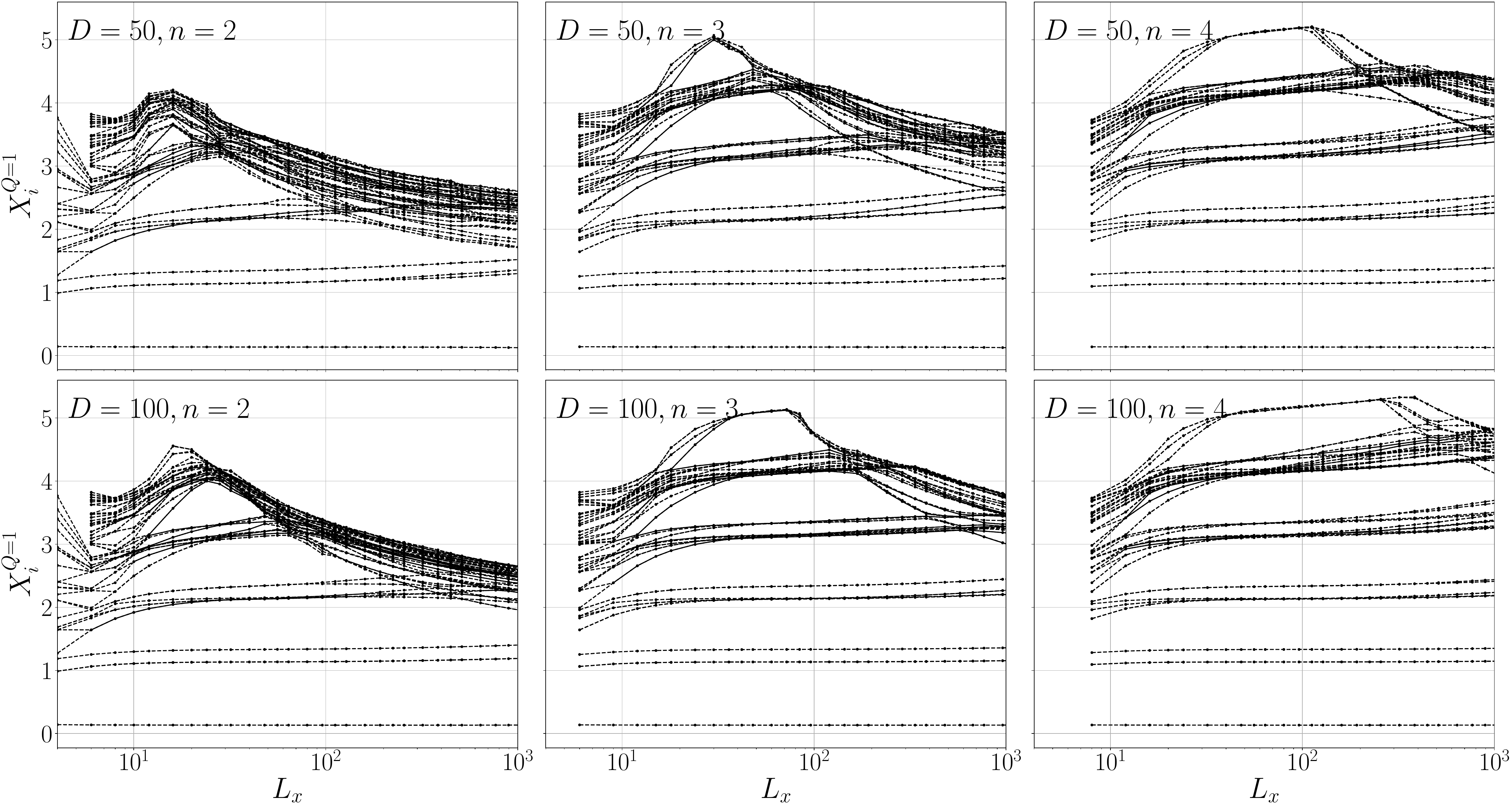}
  \caption{HOTRG results of the 3-state clock model for $X^{Q=0}_i$ (upper two rows) and $X^{Q=1}_i$ (lower two rows)   as a function of $L_x$, with $i=0\cdots 60$.
  For each $Q$, upper and lower rows correspond to bond dimensions $D=50, 100$, respectively, 
  while from left to right, the columns correspond to $n=2,3,4$, respectively.}
\label{fig:3clock_Xi_q0}
\end{center}
\end{figure}

We next apply the same framework to the 3-state clock model.
As in the Ising case, we first extract low-lying $X^{Q}_k(L_x)$ in each symmetry sector.
In Fig.~\ref{fig:3clock_Xi_q0} 
we show the HOTRG results for the lowest sixty $X^Q_k(L_x)$ in the $Q=0, +1$ sectors of the 3-state clock model.
Results for the $Q=-1$ sector are not plotted, since they are essentially identical to those for the $Q=+1$ sector.
The overall structure is qualitatively similar to that of the Ising model, but the usable scaling window is visibly narrower.
In particular, the collapse of nearby groups sets in at lower scaling dimension, which we attribute to the higher density of states.
For a fixed $D$ and $n$, we expect that the number of states one can extract from the transfer matrix
without running into the collapse problem is similar for the Ising and 3-state clock models.
The denser spectrum of the 3-state clock model therefore pushes the onset of collapse to smaller scaling dimension.

\begin{figure}[t]
  \begin{center}
   \includegraphics[width=0.95\columnwidth]{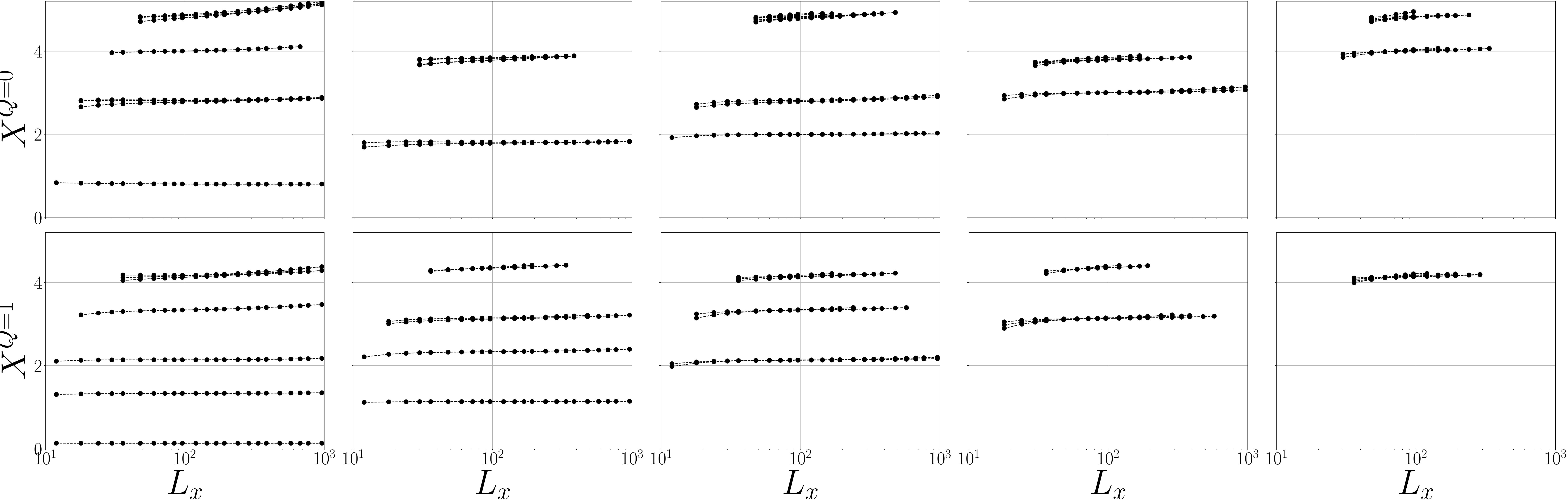}
  \caption{HOTRG results for the 3-state clock model for self-consistent spin-resolved subgroups $X^{Q,S}_k$ as a function of $L_x$ with $D=100, n=6$. 
  The first (second) row corresponds to $Q=0$ ($Q=1$), while columns from left to right correspond to $S=0$ to $S=4$.}
  \label{fig:3clock_q0_q1}
  \end{center}
\end{figure}

Next we use conformal spin to construct spin-resolved self-consistent subgroups, which are plotted in Fig.~\ref{fig:3clock_q0_q1}.
We observe that, in general, the self-consistent regime is smaller than in the Ising model.
With $D=100$ and $n=6$, we can identify spin-resolved subgroups up to $\Delta \approx 4.8$ in the $Q=0$ sector,
while the highest spin-resolved subgroup we can identify in the $Q=\pm 1$ sector has $\Delta \approx 4.33$.

These results already indicate that the 3-state clock model provides a more stringent test of the method, because operator crowding and finite-bond-dimension effects compete at smaller $L_x$.
To illustrate how the analysis adapts in this regime, we examine two representative cases before summarizing the full set of spin-resolved subgroups.
The first involves a subgroup that contains more than one CFT operator, whereas the second probes the highest-scaling-dimension subgroup that we can still resolve reliably in the $Q=1$ sector.

%

\subsubsection{$X^{Q=0}_{2, 3, 4, 5} \approx 1.8$}

\begin{figure}[t]
\begin{center}
 \includegraphics[width=0.95\columnwidth]{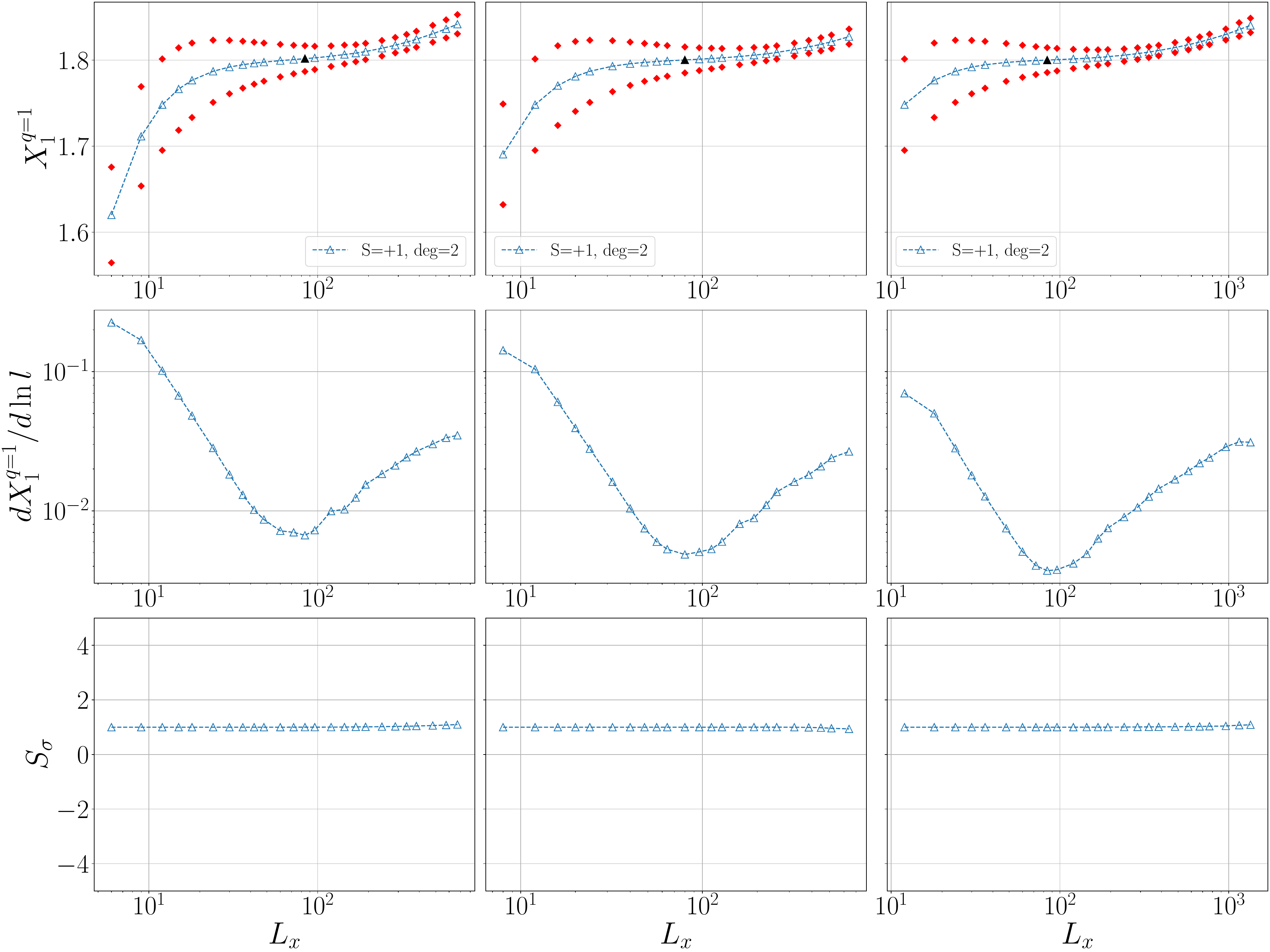}
  \caption{HOTRG results of 3-state clock model. 
                First row: $\overline{X}^{Q=0}_{S=+1}$ (dotted line) and $X^{Q=0}_{2, 3}$ as a function of $L_x$.
                Second row: $\left| \frac{d}{dl} \overline{X}^{Q=0}_{S=+1}\right|$ as a function of $L_x$.
                Third row: $S^{Q=0}_{S=+1}$ as a function of $L_x$.
                The columns correspond to different values of $n$, with $n=2,3,4$ from left to right.
                Solid symbols correspond to results at $L_x^*(D,n)$.}
\label{fig:clock_1.8}
\end{center}
\end{figure}

We begin with the second lowest group in the $Q=0$ sector, a four-state manifold with $\Delta\approx 1.8$.
Its main interest is that one spin-resolved subgroup corresponds to more than one CFT operator, while the numerical flow nevertheless remains exceptionally clean.
From the conformal spins, the group splits into two doubly degenerate branches with $S=\pm 1$.
We assign $X^{Q=0}_{2,3}$ to the $S=+1$ branch and $X^{Q=0}_{4,5}$ to the $S=-1$ branch, with states ordered by magnitude within each pair.
Figure~\ref{fig:clock_1.8} then shows the representative $S=+1$ analysis: the raw levels and their average, the derivative used to identify $L_x^*$, and the averaged spin used to delimit the self-consistent regime.
The $S=-1$ branch behaves analogously by spin-flip symmetry and is therefore omitted.
From the analysis above we identify two doubly-degenerate points in the conformal tower.
\begin{eqnarray}
  && \left( S_{\phi_1}=+1, \Delta_{\phi_1} \approx \overline{X}^{Q=0}_{S=+1} (L_x^*) \right), \text{deg}=2, \\
  && \left( S_{\phi_2}=-1, \Delta_{\phi_2} \approx  \overline{X}^{Q=0}_{S=-1} (L_x^*) \right), \text{deg}=2,
\end{eqnarray}
where $\overline{X}^{Q=0}_{S=\pm 1} (L_x^*) \approx 1.8$. 

Comparison with the exact CFT shows that the $S=+1$ branch is consistent with the degenerate pair $(\frac{7}{5}, \frac{2}{5})$ and $(\frac{2}{5}, \frac{2}{5})+(1, 0)$, while the $S=-1$ branch is consistent with $(\frac{2}{5}, \frac{7}{5})$ and $(\frac{2}{5}, \frac{2}{5})+(0, 1)$.
We therefore identify $\phi^1_1=(\frac{7}{5}, \frac{2}{5})$, $\phi^2_1=(\frac{2}{5}, \frac{2}{5})+(1, 0)$, $\phi^1_2=(\frac{2}{5}, \frac{7}{5})$, and $\phi^2_2=(\frac{2}{5}, \frac{2}{5})+(0, 1)$.
The total degeneracy is reproduced exactly, since $d(\phi^1_1)=d(\phi^2_1)=d(\phi^1_2)=d(\phi^2_2)=1$, and the relative error at $D=100$ and $n=6$ is only about $1.8\times 10^{-5}$.
This example therefore shows that the method can extract the common conformal data of multiple operators even when it cannot resolve them individually within the present framework.

\subsubsection{$X^{Q=1}_{46, \cdots, 53} \approx 4.33$}

\begin{figure}[t]
  \begin{center}
   \includegraphics[width=0.95\columnwidth]{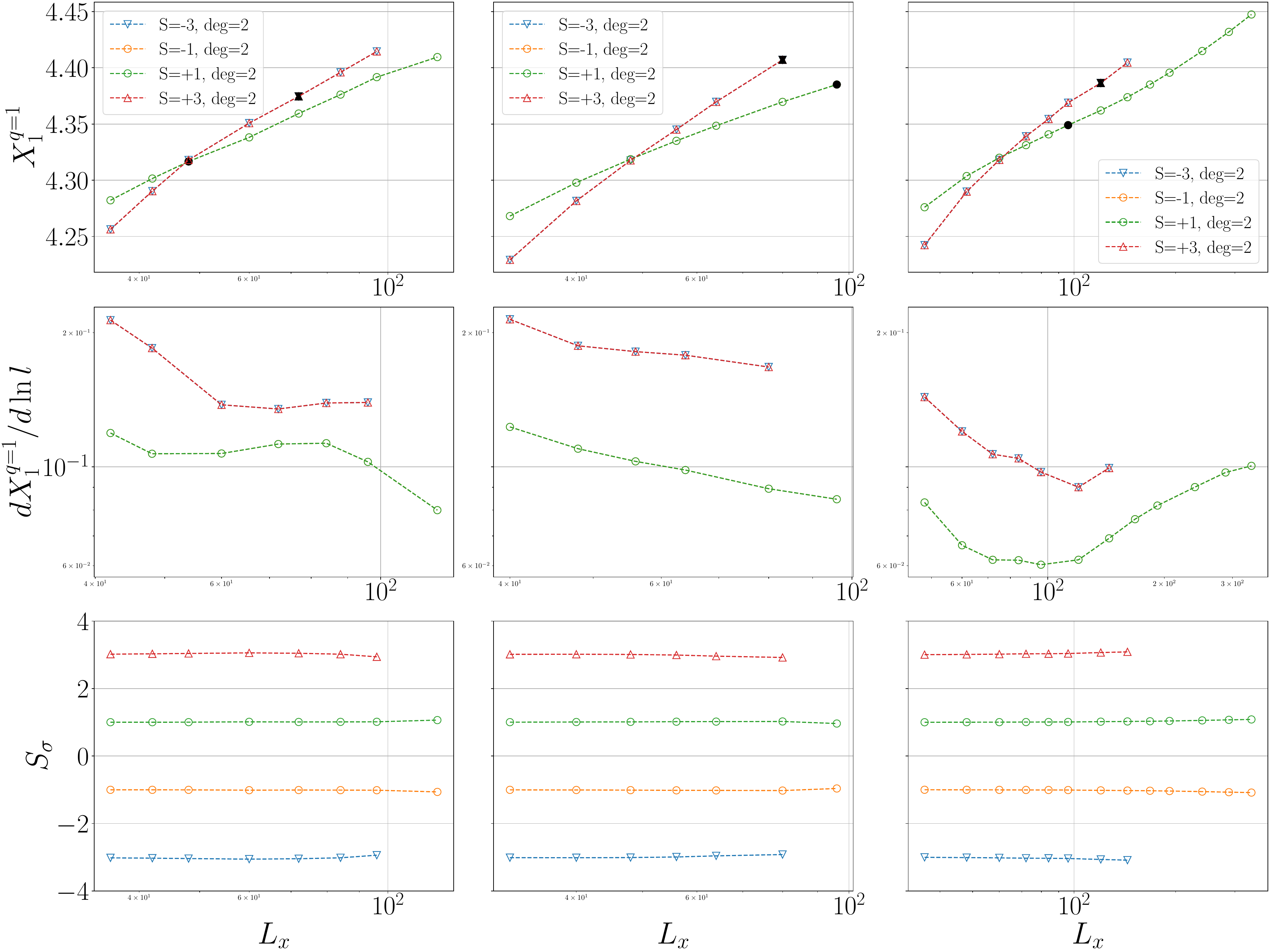}
    \caption{HOTRG results of 3-state clock model. 
                  First row: $\overline{X}^{Q=1}_{S}$ as a function of $L_x$.
                  Second row: $\left| \frac{d}{dl} \overline{X}^{Q=1}_{S} \right |$ as a function of $L_x$.
                  Third row: $S^{Q=1}_{S}$ as a function of $L_x$.
                  The columns correspond to different values of $n$, with $n=3,4,6$ from left to right.
                  Solid symbols correspond to results at $L_x^*(D,n)$.}
  \label{fig:clock_4.33}
  \end{center}
  \end{figure}

Our final case study examines the highest-scaling-dimension group that we can still resolve in the $Q=1$ sector of the 3-state clock model.
This group lies near $\Delta\approx 4.33$, contains eight states $X^{Q=1}_{46}$ through $X^{Q=1}_{53}$, and splits into four spin-resolved subgroups with $S=\pm 3$ and $\pm 1$.
Figure~\ref{fig:clock_4.33} shows the corresponding analysis based on subgroup averages for HOTRG data at $D=100$ and $n=3,4,6$.
Even at $n=6$, the self-consistent window is narrow and $\left| \frac{d} {d l} X^{Q=1}_{S} (L_x^*)\right|$ remains relatively large, making this the most demanding example in the main text.
Nevertheless, the flow still allows us to identify four points in the conformal tower and their degeneracies:
\begin{eqnarray}
  && \left( S_{\phi_1}=+3, \Delta_{\phi_1} \approx \overline{X}^{Q=1}_{S=+3} (L_x^*) \right), \text{deg}=2, \\
  && \left( S_{\phi_2}=+1, \Delta_{\phi_2} \approx \overline{X}^{Q=1}_{S=+1} (L_x^*) \right), \text{deg}=2, \\
  && \left( S_{\phi_3}=-1, \Delta_{\phi_3} \approx \overline{X}^{Q=1}_{S=-1} (L_x^*) \right), \text{deg}=2, \\
  && \left( S_{\phi_4}=-3, \Delta_{\phi_4} \approx \overline{X}^{Q=1}_{S=-3} (L_x^*) \right), \text{deg}=2,
\end{eqnarray}
where $\overline{X}^{Q=1}_{S} \approx 4.33$.

Comparison with the exact CFT identifies these four levels as
$\phi_1 = \left(\frac{2}{3},\frac{2}{3}\right)+(3,0)$,
$\phi_2 = \left(\frac{2}{3},\frac{2}{3}\right)+(2,1)$,
$\phi_3 = \left(\frac{2}{3},\frac{2}{3}\right)+(1,2)$, and
$\phi_4 = \left(\frac{2}{3},\frac{2}{3}\right)+(0,3)$.
The worst relative error is about $1.2 \times 10^{-2}$, attained for $\left(\frac{2}{3},\frac{2}{3}\right)+(3,0)$ and $\left(\frac{2}{3},\frac{2}{3}\right)+(0,3)$.
Although this is roughly one order of magnitude larger than the worst Ising error, it is still sufficiently small to support a reliable operator identification at the edge of the accessible 3-state-clock window.

Now we comment on the overall analysis of the 3-state clock model.
In Appendix~\ref{sec:conformal_data_3clock}, we present the plots of all self-consistent spin-resolved subgroups in Fig.~\ref{fig:3clock_D100n6_q0}.
Here the results are obtained with HOTRG at $D=100$ and $n=6$.
For each spin-resolved subgroup, we perform a finite-size scaling analysis 
to obtain the conformal spin, scaling dimension, degeneracy, as well as the crossover length scale.
Based on these results, we plot the conformal tower of the 3-state clock model as shown in Fig.~\ref{fig:tower_3clock}.

By comparing with the exact CFT of the 3-state clock model, we identify the corresponding CFT operator or operators for each subgroup.
We find that, in the $Q=1$ sector, each spin-resolved subgroup corresponds to a unique operator in the CFT, and the degeneracy agrees with the exact result.
On the other hand, spin-resolved subgroups in $Q=0$ may correspond to multiple distinct operators. 
In this case, we provide only one crossover length scale and one relative error.
We have checked that the total degeneracy is always in agreement.
Finally, in Tables~\ref{tab:3clock_q0-1} and \ref{tab:3clock_q1} 
we list the results of the crossover length scale $L_x^*$ and the relative error $\text{RE}_\Delta$ for $Q=0,1$ sectors respectively.
With HOTRG and $D=100$ and $n=6$, the relative error we obtained is between $1.8 \times 10^{-5}$ and $1.2 \times 10^{-2}$.
Compared with the Ising model, the relative error is overall about 10 times larger.

As in the case of the Ising model, in Tables~\ref{tab:3clock_q0-1} and \ref{tab:3clock_q1}
we also list $\tilde{L}_x^*$ and $\text{RE}_\Delta(\tilde{L}_x^*)$ for the 3-state clock model.
Here we find that the difference between $\text{RE}_\Delta(\tilde{L}_x^*)$ and
$\text{RE}_\Delta(L_x^*)$ is always small, but $\text{RE}_\Delta(\tilde{L}_x^*)$ may fall
below $\text{RE}_\Delta(L_x^*)$ even for subgroups with lower scaling dimensions.
We attribute this to the more intricate finite-size corrections in the 3-state clock model,
which stem from the larger number of irrelevant operators.
Since $\text{RE}_\Delta(\tilde{L}_x^*) \sim \text{RE}_\Delta(L_x^*) \ll 1$ for all subgroups,
our identification of $L_x^*$ remains close to optimal in practice.

\subsection{Central charge estimators \label{sec:c}}

We now compare two complementary estimators of the central charge $c$.
The first, $c_{E_0}$, is defined in Eq.~\ref{eq:c_E0} from the finite-size scaling of the ground-state energy, while the second, $c_S$, 
is defined in Eq.~\ref{eq:c_S} from the scaling of the bipartite entanglement entropy.
The usefulness of considering both is that they probe the same universal quantity through distinct observables 
and therefore provide a nontrivial internal consistency check on the finite-size analysis.

\begin{figure}[t]
  \begin{center}
   \includegraphics[width=0.95\columnwidth]{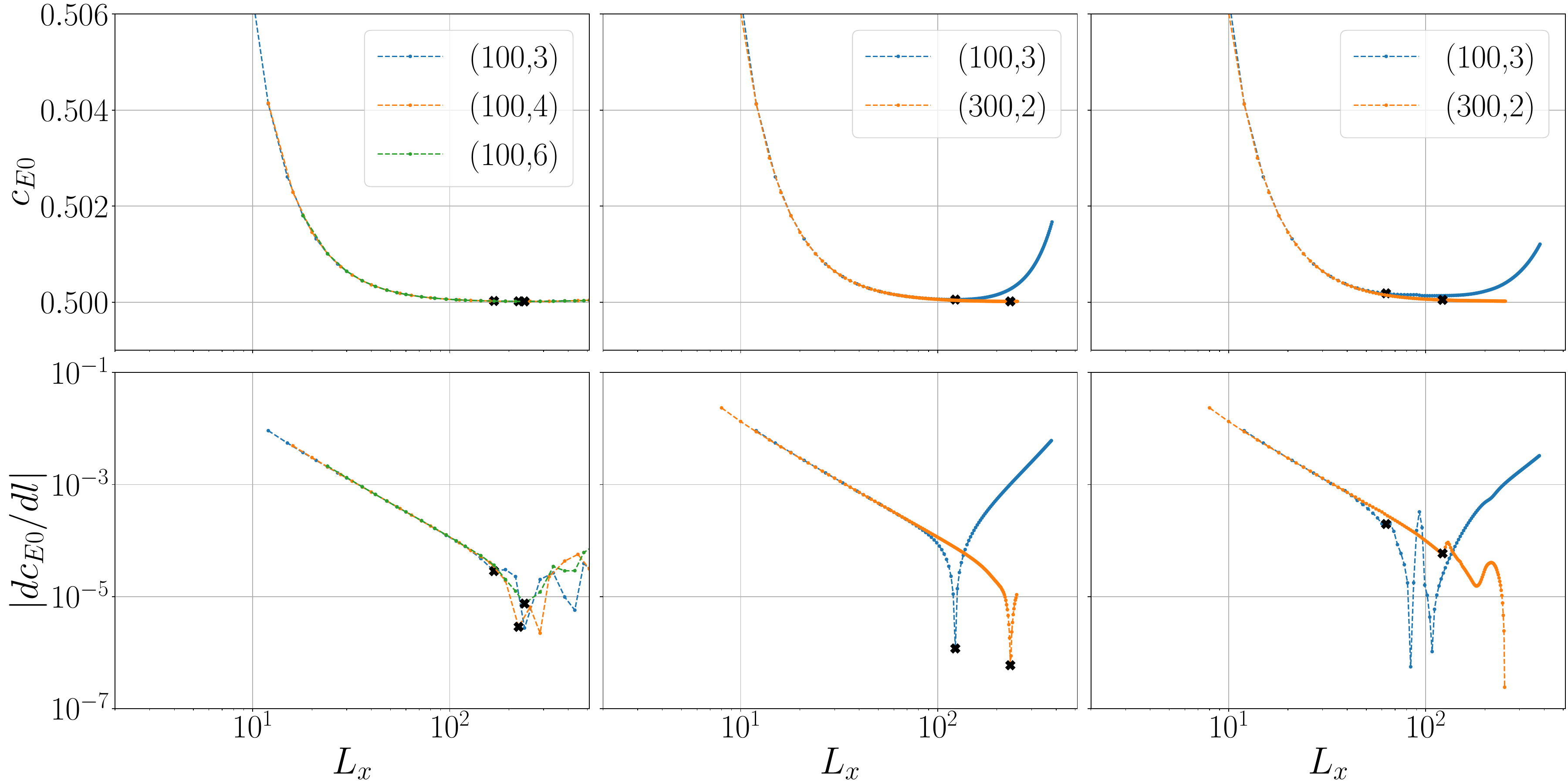}
   \includegraphics[width=0.95\columnwidth]{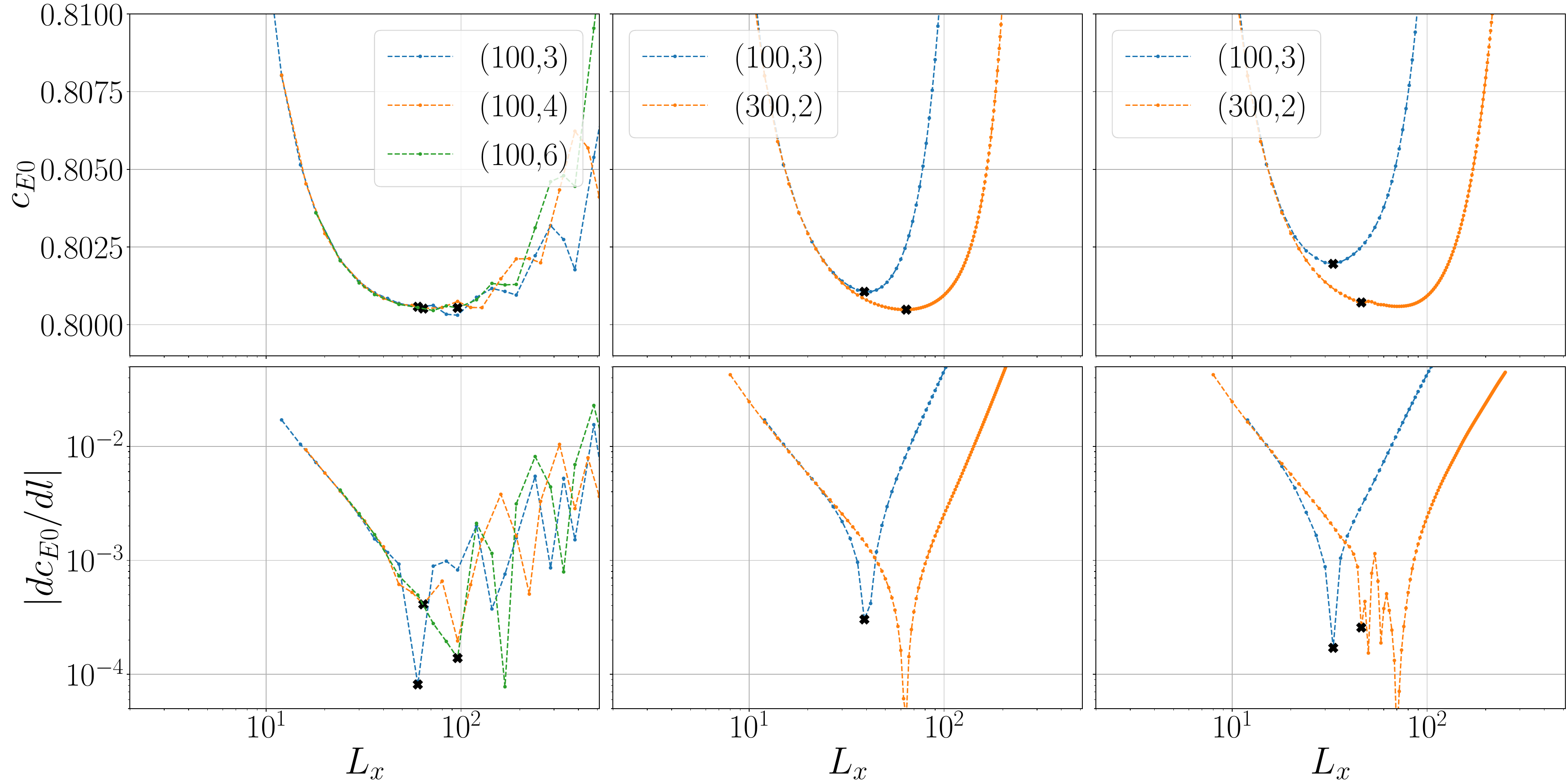}
   \caption{Central-charge estimator $c_{E_0}$ obtained from ground-state-energy scaling for the Ising and 3-state clock models.
            The first and third rows show $c_{E_0}(L_x)$, while the second and fourth rows show $\left| \frac{d c_{E_0}}{d\ln L_x} \right|$.
            From left to right, the columns correspond to HOTRG, PTMRG, and CTRG.}
  \label{fig:c_E0}
 \end{center}
\end{figure}

Figure~\ref{fig:c_E0} shows $c_{E_0}(L_x)$ and $\left| \frac{d}{dl} c_{E_0} \right|$ for the Ising and 3-state clock models.
The three columns correspond, respectively, to HOTRG, PTMRG, and CTRG.
For the method comparison we show HOTRG data at $D=100$ with $n=3,4,6$, and PTMRG/CTRG data at $D=100,n=3$ and $D=300,n=2$.
In all cases, $c_{E_0}(L_x)$ approaches the exact CFT value as $L_x$ increases up to a crossover scale.
As in the scaling-dimension analysis, $\left| \frac{d}{dl} c_{E_0} \right|$ displays an approximately universal power-law decay in the pre-crossover regime, 
followed by method-dependent behavior once finite-entanglement effects become important.
We therefore define $L_x^*$ from the minimum of $\left| \frac{d}{dl} c_{E_0} \right|$ and take $c_{E_0}(L_x^*)$ as the optimal estimate for fixed $D$ and $n$.
The relative error $\mathrm{RE}_c$ is then evaluated against the exact values $c=\frac{1}{2}$ for the Ising model and $c=\frac{4}{5}$ for the 3-state clock model.

Table~\ref{tab:c_E0} summarizes the resulting values of $L_x^*$ and $\mathrm{RE}_c$ for $c_{E_0}$.
For HOTRG with $D=100$ and $n=6$, the relative error is reduced to $3.0\times 10^{-5}$ for the Ising model and $6.8\times 10^{-4}$ for the 3-state clock model.
At fixed $D$ and $n$, PTMRG and CTRG generally yield larger errors than HOTRG, consistent with the trend reported in Ref.~\cite{Fedorovich.2025}.
The largest PTMRG/CTRG calculations considered here, at $D=300$ and $n=2$, reach an accuracy comparable to the best HOTRG results.
For the models studied in this work, HOTRG therefore provides the best overall tradeoff between accuracy and cost for central-charge extraction.

\begin{figure}[t]
  \begin{center}
   \includegraphics[width=0.95\columnwidth]{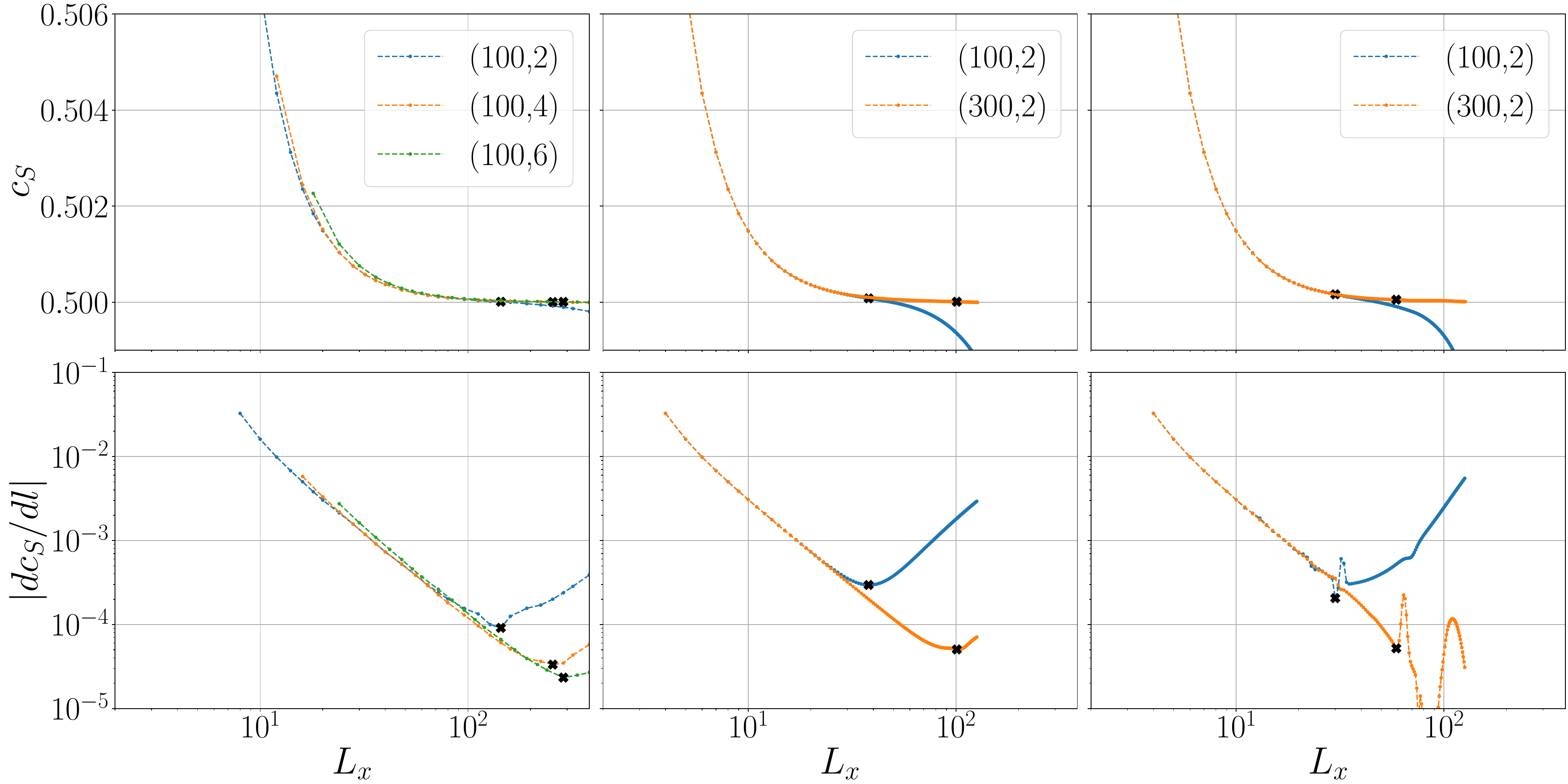}
   \includegraphics[width=0.95\columnwidth]{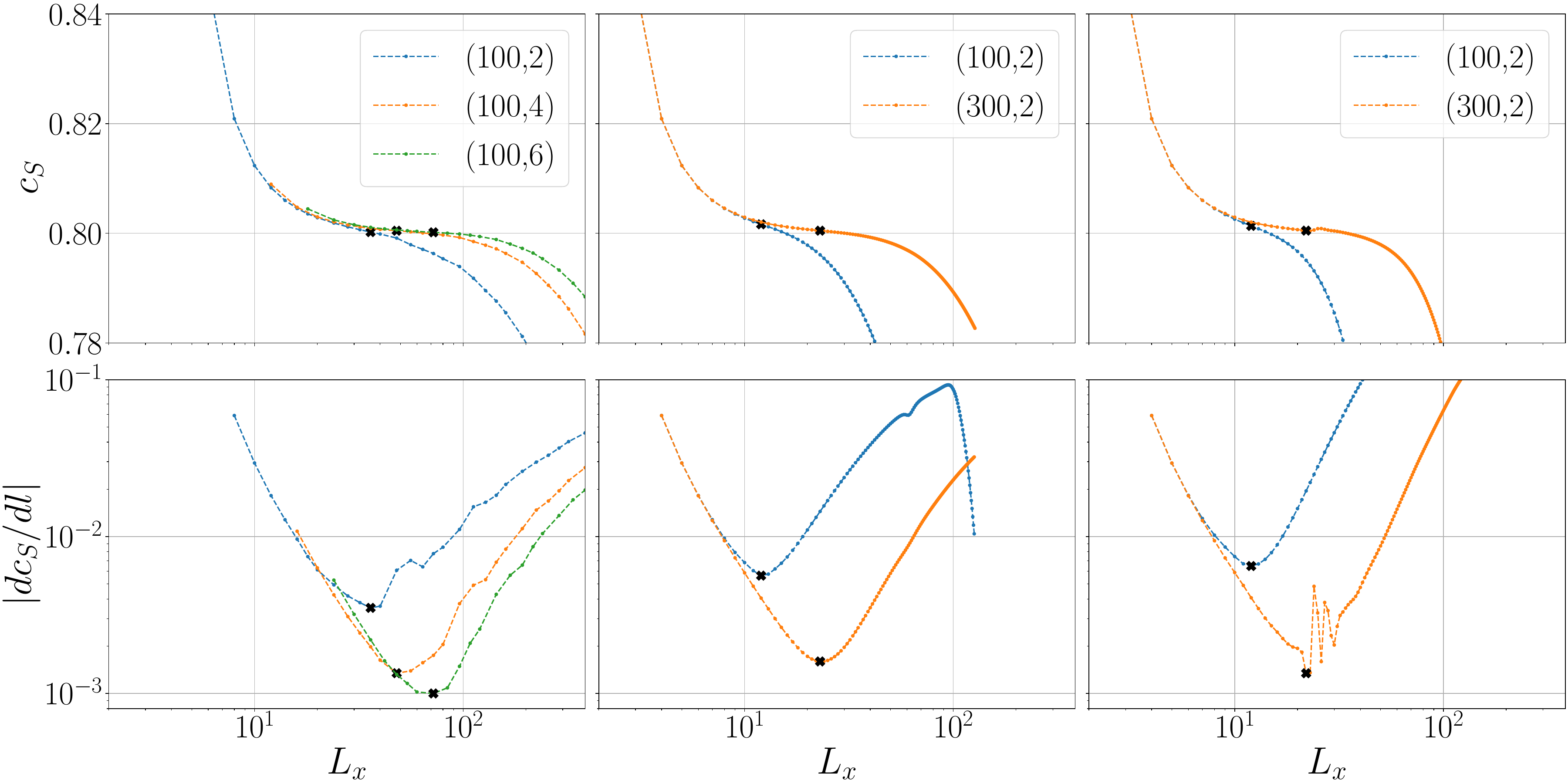}
   \caption{Central-charge estimator $c_S$ obtained from entropy scaling for the Ising and 3-state clock models.
            The first and third rows show $c_S(L_x)$, while the second and fourth rows show $\left| \frac{d c_S}{d\ln L_x} \right|$.
            From left to right, the columns correspond to HOTRG, PTMRG, and CTRG.}
  \label{fig:c_S}
  \end{center}
  \end{figure}

We next turn to the entropy-based estimator $c_S$.
Figure~\ref{fig:c_S} shows the corresponding flows of $c_S(L_x)$ and $\left| \frac{d}{dl} c_S \right|$.
The overall structure is similar to that found for $c_{E_0}$: an extended pre-crossover regime in which the data approach the CFT value, followed by a post-crossover regime in which the flow becomes scheme dependent.
Applying the same criterion to $\left| \frac{d}{dl} c_S \right|$, we extract $L_x^*$ and the associated relative error for $c_S$, with the results summarized in Table~\ref{tab:c_S}.
Quantitatively, $c_S$ is slightly more accurate than $c_{E_0}$ for all data sets considered here.
At the same time, the comparison across renormalization schemes follows the same pattern: PTMRG and CTRG typically require larger bond dimensions to reach HOTRG-level precision.
Thus, for both central-charge estimators, HOTRG again yields the best overall balance between accuracy and computational cost for the models studied in this work.

\begin{table}[b]
  \caption{\label{tab:c_E0} Relative error of the ground-state-energy estimator $c_{E_0}$ for the Ising and 3-state clock models.}
  \begin{ruledtabular}
  \begin{tabular}{ccccccc}
  method   & $D$ & n  & \multicolumn{2}{c}{Ising model}          & \multicolumn{2}{c}{3-state clock model} \\  \hline
                 &         &  & $L_x^*$ & REc                              & $L_x^*$ & REc \\  \hline
  HOTRG  & 100 & 3 & 168  & $+4.9\times 10^{-5}$    & 60 & $+7.2\times 10^{-4}$  \\ \hline 
  HOTRG  & 100 & 4 & 224 & $+3.5\times 10^{-5}$     & 64 & $+6.5\times 10^{-4}$  \\ \hline 
  HOTRG  & 100 & 6 & 240 & $+3.0\times 10^{-5}$     & 96 & $+6.8\times 10^{-4}$  \\ \hline 
  PTMRG  & 100 & 3 & 123 & $+1.1\times 10^{-4}$     & 39 & $+1.3\times 10^{-3}$  \\ \hline 
  PTMRG  & 300 & 2 & 234 & $+3.1\times 10^{-5}$     & 64 & $+6.1\times 10^{-4}$ \\ \hline 
  CTRG    & 100 & 3 & 63 &   $+3.7\times 10^{-4}$      & 33 & $+2.5\times 10^{-3}$  \\  \hline
  CTRG    & 100 & 2 & 122 & $+9.4\times 10^{-5}$      & 46 & $+9.0\times 10^{-4}$   \\  
  \end{tabular}
  \end{ruledtabular}
  \end{table}

\begin{table}[t]
\caption{\label{tab:c_S} Relative error of the entropy-based estimator $c_S$ for the Ising and 3-state clock models.}
\begin{ruledtabular}
\begin{tabular}{ccccccc}
method   & $D$ & n  & \multicolumn{2}{c}{Ising model}          & \multicolumn{2}{c}{3-state clock model} \\  \hline
               &         &  & $L_x^*$ & REc                              & $L_x^*$ & REc \\  \hline
HOTRG  & 100 & 2 & 144 & $+1.5\times 10^{-5}$    & 36 & $+2.7\times 10^{-4}$  \\ \hline 
HOTRG  & 100 & 4 & 256 & $+6.1\times 10^{-6}$    & 48 & $+5.8\times 10^{-4}$  \\ \hline 
HOTRG  & 100 & 6 & 288 & $+1.1\times 10^{-6}$    & 72 & $+2.3\times 10^{-4}$  \\ \hline 
PTMRG  & 100 & 2 & 76   & $+1.6\times 10^{-4}$    & 24 & $+2.0\times 10^{-3}$  \\ \hline 
PTMRG  & 300 & 2 & 202 & $+1.6\times 10^{-5}$    & 46 & $+6.0\times 10^{-4}$ \\ \hline 
CTRG    & 100 & 2 & 60   &  $+3.3\times 10^{-4}$    & 24 & $+1.7\times 10^{-3}$  \\  \hline
CTRG    & 300 & 2 & 118 & $+1.0\times 10^{-4}$     & 44 & $+6.1\times 10^{-4}$   \\  
\end{tabular}
\end{ruledtabular}
\end{table}

\section{Summary and outlook}

In this work, we formulated conformal-data extraction as a problem in finite-size tensor-network flow.
The central idea is that transfer-matrix spectra, together with a spin-based self-consistency criterion, 
identify a finite-size window in which universal conformal information can be extracted before bond-dimension truncation drives the flow into a finite-entanglement regime.
Within this framework, the crossover scale $L_x^*$ provides a practical and physically transparent criterion 
for selecting the data used to estimate the central charge, scaling dimensions, and conformal spins.

Benchmark calculations for the critical two-dimensional Ising and three-state clock models 
show that this framework yields accurate conformal data up to relatively high conformal levels across multiple tensor-network renormalization schemes.
The results also clarify which features of the flow are universal and which are method- and bond-dimension-dependent.
In particular, while the crossover scale depends on the observable, the renormalization scheme, 
and the bond dimension, the conformal data extracted below that scale are robust.
Within the computational range studied here, HOTRG provides the best overall performance for conformal-tower extraction, 
while PTMRG and CTRG provide useful consistency checks and complementary access to the finite-size flow.

More broadly, the framework provides a natural operational definition of entanglement scaling for classical tensor-network calculations and thereby yields a complementary estimator of the central charge.
It also opens several directions for future work, including applications to more complex classical models, systematic comparisons across a wider class of tensor-network renormalization schemes, and extensions to settings in which fixed-point-tensor approaches are difficult to control, such as non-unitary theories or models with more intricate operator spectra.

\appendix

\section{Conformal data of Ising model \label{sec:conformal_data_Ising}}

\begin{figure*}[t]
  \begin{center}
   \includegraphics[width=\textwidth]{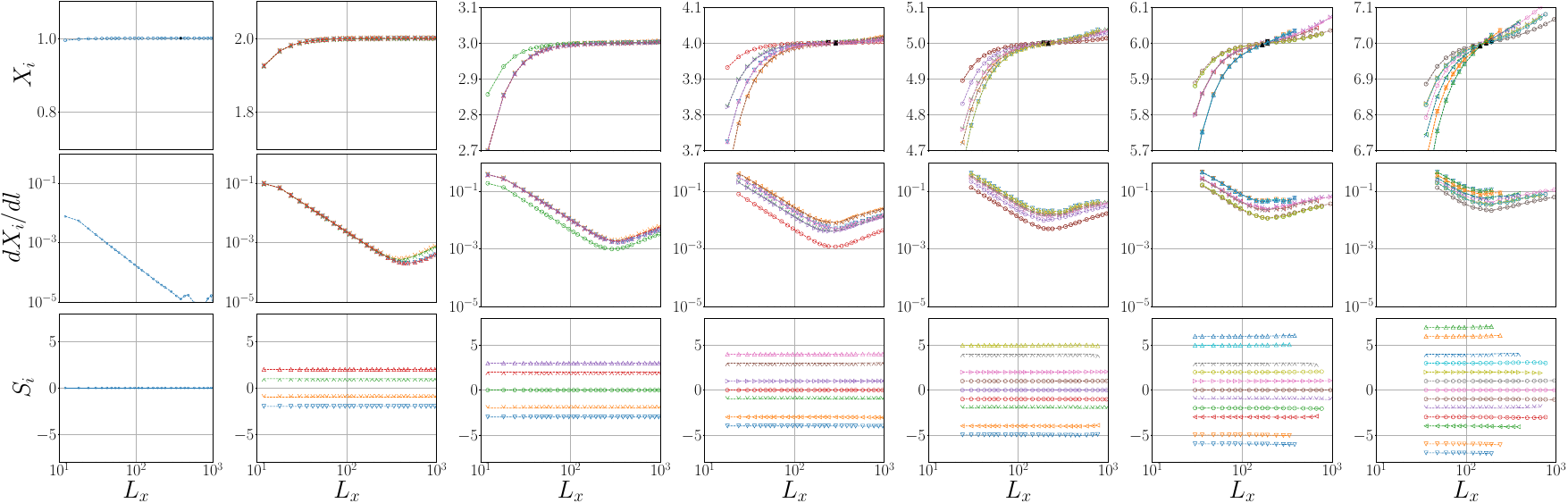}
   \includegraphics[width=\textwidth]{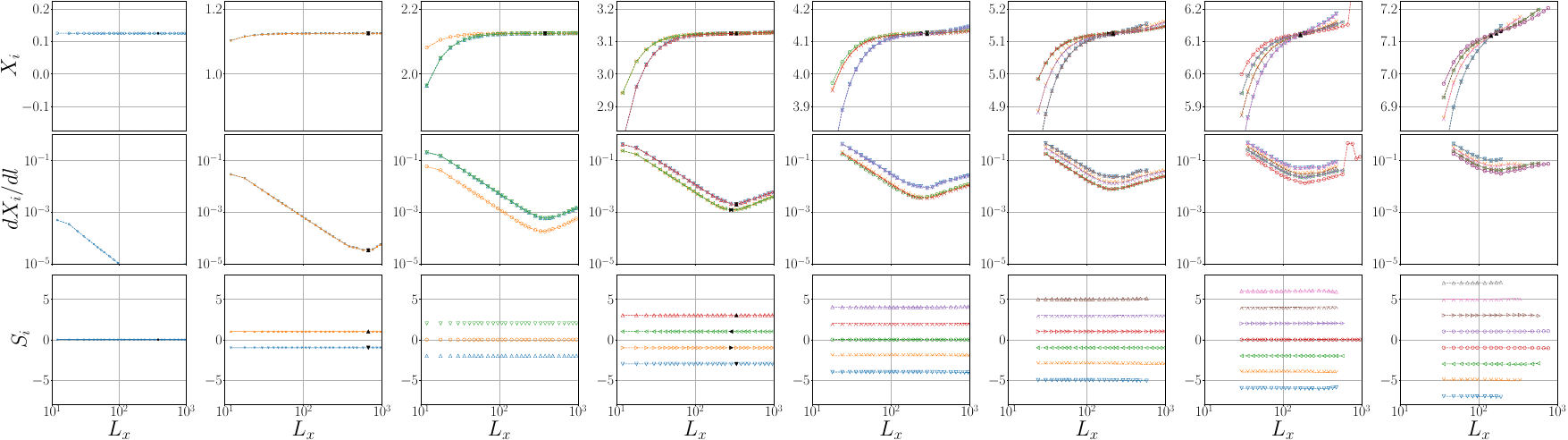}
   \caption{
   Plots of self-consistent spin-resolved subgroups obtained with HOTRG $D=100$, $n=6$ for the Ising model.
   In the 1st, 2nd, 3rd rows we plot $\overline{X}^{Q=0}_{S}$, $\left | \frac{d }{dl} \overline{X}^{Q=0}_{S} \right|$, $S^{Q=0}_S$ as a function of $L_x$.
   Corresponding plots for the $Q=1$ sector are presented in the 4th, 5th, and 6th rows.
   }
  \label{fig:Ising_D100n6_q0}
  \end{center}
\end{figure*}

In this appendix, we present exact conformal data, numerical estimates, and plots for the Ising-model results obtained with HOTRG at $D=100$ and $n=6$.
In Fig.~\ref{fig:Ising_D100n6_q0} we plot all self-consistent spin-resolved subgroups used in the analysis.
Applying our analysis, we obtain the crossover length scale $L_x^*$ and the relative error of the scaling dimension for each subgroup.
In Table~\ref{tab:Ising_q0-1},
we list all allowed operators labeled as $(h,\bar{h})+(m,\bar{m})$ in the $Q=0$ sector up to scaling dimension $7$,
while the allowed operators in the $Q=1$ sector are listed in Table~\ref{tab:Ising_q1-1} up to scaling dimension $7\frac{1}{8}$.
Each operator's degeneracy, scaling dimension, and conformal spin are also listed.
We also list the numerical results of the associated crossover length scale $L_x^*$ and the relative error of the scaling dimension $\text{RE}_\Delta$
in these tables.

\setlength{\LTleft}{0pt}
\setlength{\LTright}{0pt}
\begin{longtable}{ccccccccc}
\caption{\label{tab:Ising_q0-1} Ising model, $Q=0$ sector} \\
\hline\hline
$\left(h,\bar{h}\right)$ & $\left( I, \bar{I} \right)$ & $d \times \bar{d}$ & $\Delta$ & $S$ & $L_x^*$ & $\text{RE}_\Delta (L_x^*)$ & $\widetilde{L}_x^*$ & $\text{RE}_\Delta (\widetilde{L}_x^*)$\\
\hline\hline
\endfirsthead
\multicolumn{7}{c}{\tablename\ \thetable\ -- \textit{Continued}} \\
\hline\hline
$\left(h,\bar{h}\right)$ & $\left( I, \bar{I} \right)$ & $d \times \bar{d}$ & $\Delta$ & $S$ & $L_x^*$ & $\text{RE}_\Delta (L_x^*)$ & $\widetilde{L}_x^*$ & $\text{RE}_\Delta (\widetilde{L}_x^*)$\\
\hline\hline
\endhead
\hline
\multicolumn{7}{r}{\textit{Continued on next page}} \\
\endfoot
\hline\hline
\endlastfoot
$(0, 0)$                              &  (0,0)   &  $1 \times 1$        & 0     & 0    & -- & --  & -- & -   \\
\hline
$(\frac{1}{2}, \frac{1}{2})$   &  (0,0)   &  $1 \times 1$        &  1    & 0   & 384 & $4.0 \times 10^{-6}$ & 336 & $5.8 \times 10^{-6}$    \\
\hline
$(0, 0)$                              & (0,2)    &  $1 \times 1$        & 2    & -2    & 480 & $5.5 \times 10^{-6}$  & 432 & $1.6 \times 10^{-5}$       \\
$(0, 0)$                              &  (2,0)    & $1 \times 1$        & 2    & +2   & 480 & $5.5 \times 10^{-6}$  & 432 & $1.6 \times 10^{-5}$        \\
\hline
   $(\frac{1}{2}, \frac{1}{2})$   &  (0,1)   &  $1 \times 1$        &  2    & -1   & 384 & $2.0 \times 10^{-5}$   & 336 & $3.8 \times 10^{-5}$                 \\
   $(\frac{1}{2}, \frac{1}{2})$   &  (1,0)   &  $1 \times 1$        &  2    & +1   & 384 & $2.0 \times 10^{-5}$   & 336 & $3.8 \times 10^{-5}$                \\
\hline
   $(0, 0)$                               &  (0,3)   & $1 \times 1$          & 3     & -3    & 336 & $5.5 \times 10^{-5}$      & 288 & $1.4 \times 10^{-4}$                 \\
   $(0, 0)$                               &  (3,0)   &  $1 \times 1$         & 3     & +3    & 336 & $5.5 \times 10^{-5}$     & 288 & $1.4 \times 10^{-5}$                     \\
\hline
   $(\frac{1}{2}, \frac{1}{2})$   &  (0,2)   &  $1 \times 1$          &  3    & -2    & 336 & $3.8 \times 10^{-5}$     & 288 & $1.3 \times 10^{-4}$                 \\
   $(\frac{1}{2}, \frac{1}{2})$   &  (1,1)   &  $1 \times 1$           & 3    & 0     & 288 & $3.9 \times 10^{-5}$     & 240 & $9.9 \times 10^{-5}$              \\
   $(\frac{1}{2}, \frac{1}{2})$   &  (2,0)   &  $1 \times 1$          &  3    & +2   & 336 & $3.8 \times 10^{-5}$     & 288 & $1.3 \times 10^{-4}$                \\
\hline
  $(0, 0)$                               &  (0,4)   & $1 \times 2$           & 4    & -4  & 288 & $6.1 \times 10^{-6}$    & 240 & $2.6 \times 10^{-4}$                  \\
  $(0, 0)$                               &  (2,2)   &  $1 \times 1$          & 4    & 0   & 288 & $1.3 \times 10^{-5}$     & 240 & $6.6 \times 10^{-4}$                       \\
  $(0, 0)$                               &  (4,0)   &  $2 \times 1$          & 4     & +4  & 288 & $6.1 \times 10^{-6}$  & 240 & $2.6 \times 10^{-4}$                       \\
\hline
  $(\frac{1}{2}, \frac{1}{2})$   &  (0,3)   &  $1 \times 1$           & 4    & -3   & 288 & $1.1 \times 10^{-4}$   & 240 & $3.0 \times 10^{-4}$                       \\
  $(\frac{1}{2}, \frac{1}{2})$   &  (1,2)   &  $1 \times 1$           & 4    & -1   & 240 & $1.0 \times 10^{-4}$   & 216 & $2.1 \times 10^{-4}$                       \\
  $(\frac{1}{2}, \frac{1}{2})$   &  (2,1)   &  $1 \times 1$           & 4    & +1  & 240 & $1.0 \times 10^{-4}$    & 216 & $2.1 \times 10^{-4}$                       \\
  $(\frac{1}{2}, \frac{1}{2})$   &  (3,0)   &  $1 \times 1$           &  4   & +3  & 288 & $1.1 \times 10^{-5}$   & 240 & $3.0 \times 10^{-4}$                        \\
\hline
 $(0, 0)$                               &  (0,5)   & $1 \times 2$          & 5      & -5       & 216 & $1.9 \times 10^{-4}$   & 192 & $6.8 \times 10^{-4}$                  \\
 $(0, 0)$                               &  (2,3)   & $1 \times 1$          & 5      & -1         & 216 & $1.1 \times 10^{-4}$  & 192 & $2.3 \times 10^{-4}$                  \\
 $(0, 0)$                               &  (3,2)   & $1 \times 1$          & 5      & +1      & 216 & $1.1 \times 10^{-4}$    & 192 & $2.3 \times 10^{-4}$                \\
 $(0, 0)$                               &  (5,0)   & $2 \times 1$          & 5      & +5     & 216 & $1.9 \times 10^{-4}$     & 192 & $6.8 \times 10^{-4}$                \\
\hline
 $(\frac{1}{2}, \frac{1}{2})$   &  (0,4)   &  $1 \times 2$         & 5    & -4   & 216 & $1.5 \times 10^{-4}$  & 192 & $5.4 \times 10^{-4}$  \\
 $(\frac{1}{2}, \frac{1}{2})$   &  (1,3)   &  $1 \times 1$         & 5    & -2   & 192 & $4.0 \times 10^{-4}$  & 168 & $7.7 \times 10^{-4}$                     \\
 $(\frac{1}{2}, \frac{1}{2})$   &  (2,2)   &  $1 \times 1$         & 5    & 0    & 192 & $1.9 \times 10^{-4}$  & 168 & $4.3 \times 10^{-4}$                      \\
 $(\frac{1}{2}, \frac{1}{2})$   &  (3,1)   &  $1 \times 1$         & 5    & +2  &192 & $4.0 \times 10^{-4}$   & 168 & $7.7 \times 10^{-4}$                     \\
 $(\frac{1}{2}, \frac{1}{2})$   &  (4,0)   &  $2 \times 1$         & 5    & +4  & 216 & $1.5 \times 10^{-4}$   & 192 & $5.4 \times 10^{-4}$                     \\
\hline
$(0, 0)$                               &  (0,6)   & $1 \times 3$          & 6   & -6     & 168 & $9.4 \times 10^{-4}$    & 144 & $2.0 \times 10^{-3}$               \\
$(0, 0)$                               &  (2,4)   & $1 \times 2$          & 6   & -2     &  192& $1.7 \times 10^{-4}$   & 168 & $4.2 \times 10^{-4}$                    \\
$(0, 0)$                               &  (3,3)   & $1 \times 1$          & 6   & 0      & 192 & $2.4 \times 10^{-5}$    & 168 & $2.7 \times 10^{-4}$                          \\
$(0, 0)$                               &  (4,2)   & $2 \times 1$          & 6   & +2    & 192 & $1.7 \times 10^{-4}$    & 168 & $4.2 \times 10^{-4}$                     \\
$(0, 0)$                               &  (6,0)   & $3 \times 1$          & 6   & +6    & 168 & $9.4 \times 10^{-4}$    & 144 & $2.0 \times 10^{-3}$                    \\
\hline
$(\frac{1}{2}, \frac{1}{2})$   &  (0,5)   &  $1 \times 2$         & 6    & -5    & 168 & $8.3 \times 10^{-4}$    & 144 & $2.0 \times 10^{-3}$                     \\
$(\frac{1}{2}, \frac{1}{2})$   &  (1,4)   &  $1 \times 2$         & 6    & -3   & 192 & $8.8 \times 10^{-4}$    & 168 & $5.1 \times 10^{-4}$                          \\
$(\frac{1}{2}, \frac{1}{2})$   &  (2,3)   &  $1 \times 1$         & 6    & -1  & 192& $3.3 \times 10^{-4}$     & 168 & $1.2 \times 10^{-4}$                   \\
$(\frac{1}{2}, \frac{1}{2})$   &  (3,2)   &  $1 \times 1$         & 6    & +1 & 168 & $3.3 \times 10^{-4}$    & 168 & $1.2 \times 10^{-4}$                    \\
$(\frac{1}{2}, \frac{1}{2})$   &  (4,1)   &  $2 \times 1$         & 6    & +3 & 192 & $8.8 \times 10^{-4}$    & 168 & $5.1 \times 10^{-4}$                    \\
$(\frac{1}{2}, \frac{1}{2})$   &  (5,0)   &  $2 \times 1$         & 6   & +5 & 168 & $8.3 \times 10^{-4}$     & 144 & $2.0 \times 10^{-3}$                  \\
\hline
$(0, 0)$                               &  (0,7)   & $1 \times 3$          & 7    & -7   & 144 & $1.3 \times 10^{-3}$  & 120 & $4.3 \times 10^{-3}$  \\
$(0, 0)$                               &  (2,5)   & $1 \times 2$          & 7    & -3    & 168 & $1.8 \times 10^{-4}$   & 144 & $9.5 \times 10^{-4}$                   \\
$(0, 0)$                               &  (3,4)   & $1 \times 2$          & 7    & -1      & 192 & $3.5 \times 10^{-4}$   & 168 & $2.6 \times 10^{-5}$                      \\
$(0, 0)$                               &  (4,3)   & $2 \times 1$          & 7    & +1      & 192 & $3.5 \times 10^{-4}$  & 168 & $2.6 \times 10^{-5}$                    \\
$(0, 0)$                               &  (5,2)   & $2 \times 1$          & 7    &+3   & 168 & $1.8 \times 10^{-4}$   & 144 & $9.5 \times 10^{-4}$                    \\
$(0, 0)$                               &  (7,0)   & $3 \times 1$          & 7    & +7  & 144 & $1.3 \times 10^{-3}$  & 120 & $4.3 \times 10^{-3}$                        \\
\hline
$(\frac{1}{2}, \frac{1}{2})$   &  (0,6)   &  $1 \times 3$         & 7    & -6   & 144 & $1.2 \times 10^{-3}$  & 120 & $3.7 \times 10^{-3}$                     \\
$(\frac{1}{2}, \frac{1}{2})$   &  (1,5)   &  $1 \times 2$         & 7    & -4     & 168 & $2.3 \times 10^{-4}$ & 144 & $1.0 \times 10^{-3}$                  \\
$(\frac{1}{2}, \frac{1}{2})$   &  (2,4)   &  $1 \times 2$         & 7    & -2     & 192 & $9.4 \times 10^{-4}$ & 168 & $3.6 \times 10^{-4}$                   \\
$(\frac{1}{2}, \frac{1}{2})$   &  (3,3)   &  $1 \times 1$         & 7    & +0      & 120 & $1.5 \times 10^{-3}$ & 96 & $3.7 \times 10^{-3}$                     \\
$(\frac{1}{2}, \frac{1}{2})$   &  (4,2)   &  $2 \times 1$         & 7     & +2  & 192 & $9.4 \times 10^{-4}$    & 168 & $3.6 \times 10^{-4}$                     \\
$(\frac{1}{2}, \frac{1}{2})$   &  (5,1)   &  $2 \times 1$         & 7    & +4   & 168 & $2.3 \times 10^{-4}$  & 144 & $1.0 \times 10^{-3}$                  \\
$(\frac{1}{2}, \frac{1}{2})$   &  (6,0)   &  $3 \times 1$         & 7     & +6  & 144 & $1.2 \times 10^{-3}$  & 120 & $3.7 \times 10^{-3}$                    \\
\end{longtable}

\setlength{\LTleft}{0pt}
\setlength{\LTright}{0pt}
\begin{longtable}{ccccccccc}
\caption{\label{tab:Ising_q1-1} Ising model, $Q=1$ sector} \\
\hline\hline
$\left(h,\bar{h}\right)$ & $\left( I, \bar{I} \right)$ & $d \times \bar{d}$ & $\Delta$ & $S$ & $L_x^*$ & $\text{RE}_\Delta (L_x^*)$ & $\widetilde{L}_x^*$ & $\text{RE}_\Delta (\widetilde{L}_x^*)$ \\
\hline\hline
\endfirsthead
\multicolumn{7}{c}{\tablename\ \thetable\ -- \textit{Continued}} \\
\hline\hline
$\left(h,\bar{h}\right)$ & $\left( I, \bar{I} \right)$ & $d \times \bar{d}$ & $\Delta$ & $S$ & $L_x^*$ & $\text{RE}_\Delta (L_x^*)$ & $\widetilde{L}_x^*$ & $\text{RE}_\Delta (\widetilde{L}_x^*)$ \\
\hline\hline
\endhead
\hline
\multicolumn{7}{r}{\textit{Continued on next page}} \\
\endfoot
\hline\hline
\endlastfoot
$(\frac{1}{16}, \frac{1}{16})$ & (0,0)    &  $1 \times 1$   & $\frac{1}{8}$   &   0   & 384 & $3.7 \times 10^{-6}$   & 336 & $5.8 \times 10^{-6}$                           \\
\hline
$(\frac{1}{16}, \frac{1}{16})$  & (0,1)    &  $1 \times 1$  & $1\frac{1}{8}$  &  -1   & 672 & $2.2 \times 10^{-6}$    & 572& $2.4 \times 10^{-5}$                \\
$(\frac{1}{16}, \frac{1}{16})$  & (1,0)    &  $1 \times 1$  & $1\frac{1}{8}$  &  +1   & 672  & $2.2 \times 10^{-6}$   & 572& $2.4 \times 10^{-5}$                \\
\hline
$(\frac{1}{16}, \frac{1}{16})$  & (0,2)    &  $1 \times 1$  & $2\frac{1}{8}$  &  -1  & 384 & $4.0 \times 10^{-5}$  & 336 & $8.7 \times 10^{-5}$                    \\
$(\frac{1}{16}, \frac{1}{16})$  & (1,1)    &  $1 \times 1$  & $2\frac{1}{8}$   &  0  & 384 & $3.9 \times 10^{-6}$   & 336 & $1.5 \times 10^{-5}$                     \\
$(\frac{1}{16}, \frac{1}{16})$  & (2,0)    &  $1 \times 1$ & $2\frac{1}{8}$  &  +1  & 384 & $4.0 \times 10^{-5}$  & 336 & $8.7 \times 10^{-5}$                      \\
\hline
$(\frac{1}{16}, \frac{1}{16})$  & (0,3)    &  $1 \times 2$  & $3\frac{1}{8}$  &  -3   & 336  & $3.9 \times 10^{-5}$   & 288 & $1.3 \times 10^{-4}$               \\
$(\frac{1}{16}, \frac{1}{16})$  & (1,2)    &  $1 \times 1$  & $3\frac{1}{8}$  &  -1   & 288  & $5.5 \times 10^{-5}$    & 240 & $1.3 \times 10^{-4}$               \\
$(\frac{1}{16}, \frac{1}{16})$  & (2,1)    &  $1 \times 1$  & $3\frac{1}{8}$   &  +1  & 288  & $5.5 \times 10^{-5}$    & 248 & $1.3 \times 10^{-4}$                \\
$(\frac{1}{16}, \frac{1}{16})$  & (3,0)    &  $2 \times 1$ & $3\frac{1}{8}$    &  +3  & 366   & $3.9 \times 10^{-5}$   & 288 & $1.3 \times 10^{-4}$               \\
\hline
$(\frac{1}{16}, \frac{1}{16})$  & (0,4)    &  $1 \times 2$  & $4\frac{1}{8}$  &  -4   & 288 &  $2.2 \times 10^{-4}$     & 240 & $2.4 \times 10^{-4}$                \\
$(\frac{1}{16}, \frac{1}{16})$  & (1,3)    &  $1 \times 2$  & $4\frac{1}{8}$  &  -2    & 288 &  $2.9 \times 10^{-5}$    & 240 & $1.3 \times 10^{-4}$            \\
$(\frac{1}{16}, \frac{1}{16})$  & (2,2)    &  $1 \times 1$  & $4\frac{1}{8}$ &  0      & 240 &  $4.4 \times 10^{-5}$    & 216 & $1.4 \times 10^{-4}$           \\
$(\frac{1}{16}, \frac{1}{16})$  & (3,1)    &  $2 \times 1$  & $4\frac{1}{8}$ &  +2    & 288 & $2.9 \times 10^{-5}$     & 240 & $1.3 \times 10^{-4}$             \\
$(\frac{1}{16}, \frac{1}{16})$  & (4,0)    &  $2 \times 1$ & $4\frac{1}{8}$  &  +4   & 288 &  $2.2 \times 10^{-4}$     & 240 & $2.4 \times 10^{-4}$                \\
\hline
$(\frac{1}{16}, \frac{1}{16})$  & (0,5)    &  $1 \times 3$  & $5\frac{1}{8}$   &  -5  & 216 & $2.4 \times 10^{-5}$  & 192 & $5.8 \times 10^{-4}$                    \\
$(\frac{1}{16}, \frac{1}{16})$  & (1,4)    &  $1 \times 2$  & $5\frac{1}{8}$   &  -3  & 216 & $1.3 \times 10^{-4}$  & 192 & $4.5 \times 10^{-4}$                         \\
$(\frac{1}{16}, \frac{1}{16})$  & (2,3)    &  $1 \times 2$  & $5\frac{1}{8}$  &  -1   & 192 & $2.0 \times 10^{-4}$  & 168 & $4.0 \times 10^{-4}$                  \\
$(\frac{1}{16}, \frac{1}{16})$  & (3,2)    &  $2 \times 1$  & $5\frac{1}{8}$  &  +1 &  192 &$2.0 \times 10^{-4}$   & 168 & $4.0 \times 10^{-4}$                  \\
$(\frac{1}{16}, \frac{1}{16})$  & (4,1)    &  $2 \times 1$  & $5\frac{1}{8}$   &  +3  & 216 & $1.3 \times 10^{-4}$ & 192 & $4.5 \times 10^{-4}$                 \\
$(\frac{1}{16}, \frac{1}{16})$  & (5,0)    &  $3 \times 1$ & $5\frac{1}{8}$   &  +5 &  216 & $2.4 \times 10^{-5}$  & 192 & $5.8 \times 10^{-4}$                    \\
\hline
$(\frac{1}{16}, \frac{1}{16})$  & (0,6)    &  $1 \times 4$  & $6\frac{1}{8}$  &  -6     & 168 & $6.7 \times 10^{-4}$  & 144 & $1.9 \times 10^{-3}$                \\
$(\frac{1}{16}, \frac{1}{16})$  & (1,5)    &  $1 \times 3$  & $6\frac{1}{8}$  &  -4     & 168 & $6.9 \times 10^{-4}$  & 144 & $1.5 \times 10^{-3}$               \\
$(\frac{1}{16}, \frac{1}{16})$  & (2,4)    &  $1 \times 2$  & $6\frac{1}{8}$   &  -2     & 192 & $1.4 \times 10^{-4}$ & 168 & $3.1 \times 10^{-4}$               \\
$(\frac{1}{16}, \frac{1}{16})$  & (3,3)    &  $2 \times 2$  & $6\frac{1}{8}$  &  0       & 192 & $2.8 \times 10^{-5}$ & 168 & $2.4 \times 10^{-4}$               \\
$(\frac{1}{16}, \frac{1}{16})$  & (4,2)    &  $2 \times 1$  & $6\frac{1}{8}$  &  +2     & 192 & $1.4 \times 10^{-4}$ & 168 & $3.1 \times 10^{-4}$                \\
$(\frac{1}{16}, \frac{1}{16})$  & (5,1)    &  $3 \times 1$  & $6\frac{1}{8}$   &  +4    & 168 & $6.9 \times 10^{-4}$ & 144 & $1.5 \times 10^{-3}$               \\
$(\frac{1}{16}, \frac{1}{16})$  & (6,0)    &  $4 \times 1$ & $6\frac{1}{8}$  &  +6    & 168 & $6.7 \times 10^{-4}$  & 144 & $1.9 \times 10^{-3}$                \\
\hline
$(\frac{1}{16}, \frac{1}{16})$  & (0,7)    &  $1 \times 5$  & $7\frac{1}{8}$  &  -7   & 144 & $1.1 \times 10^{-3}$  & 120 & $3.8 \times 10^{-3}$                     \\
$(\frac{1}{16}, \frac{1}{16})$  & (1,6)    &  $1 \times 4$  & $7\frac{1}{8}$  &  -5   & 168 & $1.8 \times 10^{-4}$   & 144 & $1.1 \times 10^{-3}$               \\
$(\frac{1}{16}, \frac{1}{16})$  & (2,5)    &  $1 \times 3$  & $7\frac{1}{8}$  &  -3    & 192 & $9.7 \times 10^{-4}$  & 168 & $2.1 \times 10^{-4}$               \\
$(\frac{1}{16}, \frac{1}{16})$  & (3,4)    &  $2 \times 2$  & $7\frac{1}{8}$  &  -1     & 192 & $8.3 \times 10^{-4}$  & 168 & $3.0 \times 10^{-4}$                \\
$(\frac{1}{16}, \frac{1}{16})$  & (4,3)    &  $2 \times 2$  & $7\frac{1}{8}$  &  +1     & 192 & $8.3 \times 10^{-4}$  & 168 & $3.0 \times 10^{-4}$              \\
$(\frac{1}{16}, \frac{1}{16})$  & (5,2)    &  $3 \times 1$  & $7\frac{1}{8}$  &  +3    & 192 & $9.7 \times 10^{-4}$   & 168 & $2.1 \times 10^{-4}$             \\
$(\frac{1}{16}, \frac{1}{16})$  & (6,1)    &  $4 \times 1$  & $7\frac{1}{8}$  &  +5   & 168 & $1.8 \times 10^{-4}$  & 144 & $1.1 \times 10^{-3}$                \\
$(\frac{1}{16}, \frac{1}{16})$  & (7,0)    &  $5 \times 1$ & $7\frac{1}{8}$  &  +7      & 144 & $1.1 \times 10^{-3}$  & 120 & $3.8 \times 10^{-3}$               \\
\end{longtable}

\section{Conformal data of 3-state clock model \label{sec:conformal_data_3clock}}    

\begin{figure*}[h]
  \begin{center}
   \includegraphics[width=\textwidth]{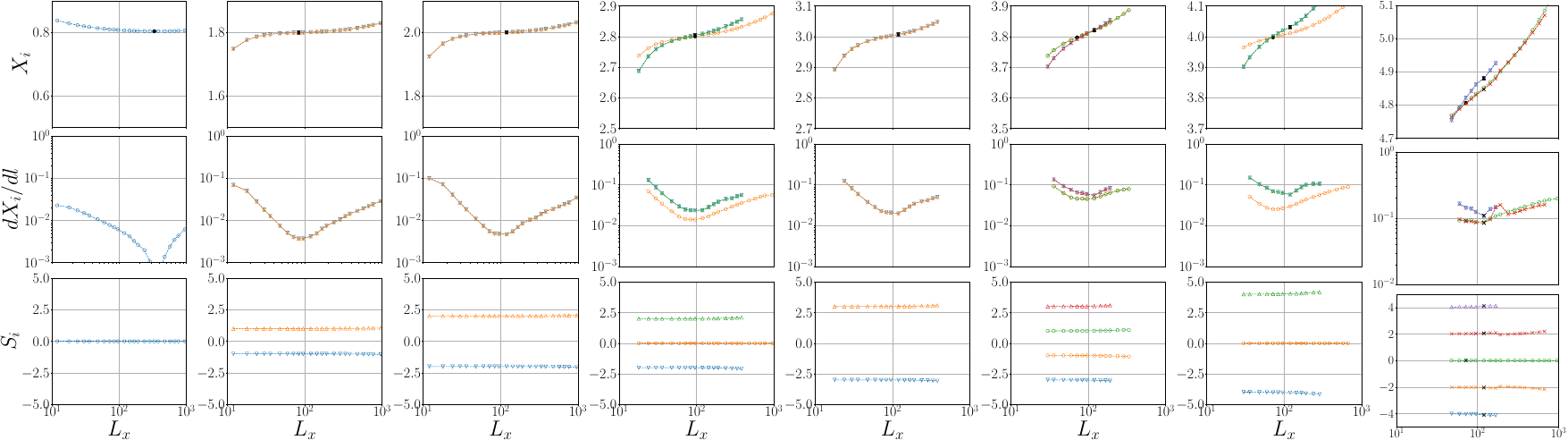}
   \includegraphics[width=\textwidth]{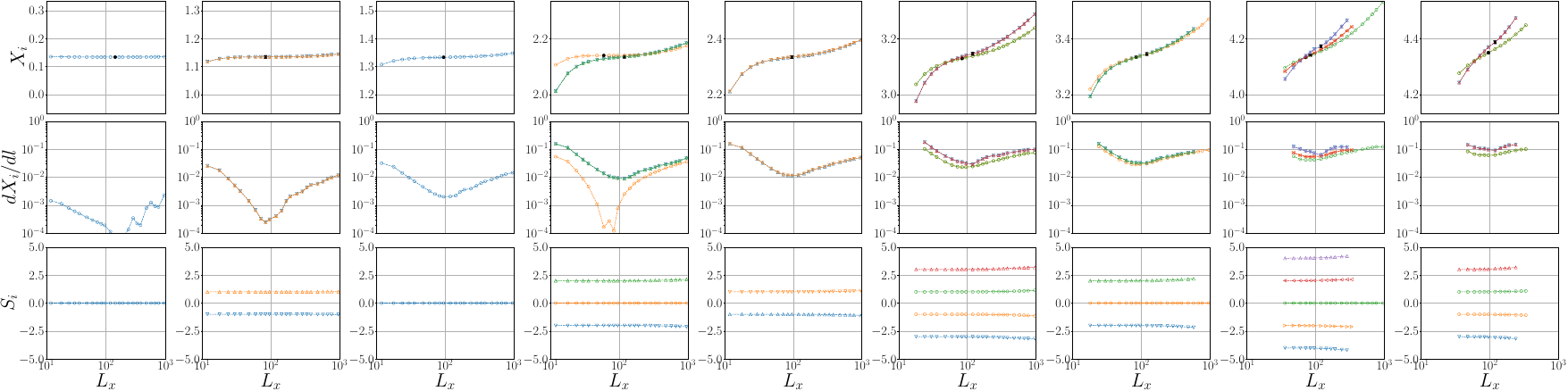}
   \caption{
   Plots of self-consistent spin-resolved subgroups obtained with HOTRG $D=100$, $n=6$ for the 3-state clock model.
   In the 1st, 2nd, 3rd rows we plot $\overline{X}^{Q=0}_{S}$, $\left | \frac{d }{dl} \overline{X}^{Q=0}_{S} \right|$, $S^{Q=0}_S$ as a function of $L_x$.
   Corresponding plots for the $Q=1$ sector are presented in the 4th, 5th, and 6th rows.
   }
  \label{fig:3clock_D100n6_q0}
  \end{center}
\end{figure*}

In this appendix, we present exact conformal data, numerical estimates, and plots for the 3-state-clock-model results obtained with HOTRG at $D=100$ and $n=6$.
In Fig.~\ref{fig:3clock_D100n6_q0} we plot all self-consistent spin-resolved subgroups used in the analysis.
Applying our analysis, we obtain the crossover length scale $L_x^*$ and the relative error of the scaling dimension for each subgroup.
For the 3-state clock model, we list allowed operators in the $Q=0$ sector up to scaling dimension $4$
in Table~\ref{tab:3clock_q0-1}, while in Table~\ref{tab:3clock_q1} 
we list allowed operators in the $Q=+1$ sector up to scaling dimension $4\frac{1}{3}$. 
Since $Q=\pm 1$ sectors have exactly the same structure, we do not separately list operators in the $Q=-1$ sector.
In these two tables, we also list the numerical results of the associated crossover length scale $L_x^*$ 
and the relative error of the scaling dimension $\text{RE}_\Delta$.
Note that multiple operators might have exactly the same scaling dimension and conformal spin.
As discussed in the main text, we report one crossover length scale $L_x^*$ and one scaling dimension 
for each unique pair of conformal spin and scaling dimension.

\setlength{\LTleft}{0pt}
\setlength{\LTright}{0pt}
\begin{longtable}{ccccccccc}
\caption{\label{tab:3clock_q0-1} 3-state clock model, $Q=0$ sector} \\
\hline\hline
$\left(h,\bar{h}\right)$ & $\left( I, \bar{I} \right)$ & $d \times \bar{d}$ & $\Delta$ & $S$ & $L_x^*$ & $\text{RE}_\Delta (L_x^*)$ & $\widetilde{L}_x^*$ & $\text{RE}_\Delta (\widetilde{L}_x^*)$ \\
\hline\hline
\endfirsthead
\multicolumn{7}{c}{\tablename\ \thetable\ -- \textit{Continued}} \\
\hline\hline
$\left(h,\bar{h}\right)$ & $\left( I, \bar{I} \right)$ & $d \times \bar{d}$ & $\Delta$ & $S$ & $L_x^*$ & $\text{RE}_\Delta (L_x^*)$ & $\widetilde{L}_x^*$ & $\text{RE}_\Delta (\widetilde{L}_x^*)$ \\
\hline\hline
\endhead
\hline
\multicolumn{7}{r}{\textit{Continued on next page}} \\
\endfoot
\hline\hline
\endlastfoot
$(0, 0)$                               &  (0,0)   &  $1 \times 1$          & $0$                   & 0   & -- & -- & -- & --\\
\hline
$(\frac{2}{5}, \frac{2}{5})$   &  (0,0)   &  $1 \times 1$          &  $\frac{4}{5}$    & 0  & 336 & $5.1 \times 10^{-3}$     & 288 & $5.2 \times 10^{-3}$                        \\
\hline
$(\frac{2}{5}, \frac{2}{5})$   &  (0,1)   &  $1 \times 1$         &  $1\frac{4}{5}$    & -1        & 84 & $1.8 \times 10^{-5}$ & 72 & $3.4 \times 10^{-4}$                \\
$(\frac{2}{5}, \frac{7}{5})$   &  (0,0)   &  $1 \times 1$          &  $1\frac{4}{5}$    & -1        & --               & -- & --               & --\\
$(\frac{2}{5}, \frac{2}{5})$   &  (1,0)   &  $1 \times 1$          &  $1\frac{4}{5}$    & +1          & 84 & $1.8 \times 10^{-5}$  & 72 & $3.4 \times 10^{-4}$             \\
$(\frac{7}{5}, \frac{2}{5})$   &  (0,0)   &  $1 \times 1$          &  $1\frac{4}{5}$     & +1        & --              & -- & --               & -- \\
\hline
$(0, 0)$                              & (0,2)    &  $1 \times 1$          & $2$                 & -2      & 120 & $3.2 \times 10^{-4}$     & 96 & $1.8 \times 10^{-4}$                  \\
$(0, 0)$                              &  (2,0)    & $1 \times 1$          & $2$                 & +2     & 120 & $3.2 \times 10^{-4}$     & 96 & $1.8 \times 10^{-4}$              \\
\hline
$(\frac{2}{5}, \frac{2}{5})$   &  (0,2)   &  $1 \times 1$           &  $2\frac{4}{5}$    & -2    & 96 & $1.5 \times 10^{-3}$    & 84 & $3.2 \times 10^{-4}$                       \\
$(\frac{2}{5}, \frac{7}{5})$   &  (0,1)   &  $1 \times 1$           &  $2\frac{4}{5}$    & -2     & --                     & --  & --                     & --  \\ 
$(\frac{7}{5}, \frac{7}{5})$  &  (0,0)   &  $1 \times 1$           & $2\frac{4}{5}$     & 0    & 96 & $1.7 \times 10^{-4}$  & 96 & $4.8 \times 10^{-4}$                      \\
$(\frac{7}{5}, \frac{2}{5})$    &  (0,1)   &  $1 \times 1$          &  $2\frac{4}{5}$    & 0      & --                    & -- & --                    & --\\
$(\frac{2}{5}, \frac{2}{5})$   &  (1,1)   &  $1 \times 1$           &  $2\frac{4}{5}$    & 0     & --                    & --  & --                    & --\\
$(\frac{2}{5}, \frac{7}{5})$   &  (1,0)   &  $1 \times 1$           &  $2\frac{4}{5}$    & 0      & --                    & -- & --                    & --\\
$(\frac{2}{5}, \frac{2}{5})$   &  (2,0)   &  $1 \times 1$           &  $2\frac{4}{5}$    & +2   & 96 & $1.5 \times 10^{-3}$    & 84 & $3.2 \times 10^{-4}$                    \\
$(\frac{7}{5}, \frac{2}{5})$    &  (1,0)   &  $1 \times 1$          &  $2\frac{4}{5}$     & +2    & --                      & -- & --                      & -- \\
\hline
$(0,3)$                              &   (0,0)   &  $1 \times 1$          & 3                           & -3      & 120 & $2.2 \times 10^{-3}$     & 96 & $9.0 \times 10^{-4}$                  \\
$(0, 0)$                             &  (0,3)   &  $1 \times 1$           & 3                           & -3        & --                   & -- & --                   & --\\
$(3,0)$                              &   (0,0)   &  $1 \times 1$          & 3                           & +3      & 120 & $2.2 \times 10^{-3}$    & 96 & $9.0 \times 10^{-4}$                    \\
$(0, 0)$                              &  (3,0)   &  $1 \times 1$          & 3                           & +3        & --                  & -- & --                   & --\\
\hline
$(\frac{2}{5}, \frac{7}{5})$   &  (0,2)   &  $1 \times 2$       &  $3\frac{4}{5}$    & -3         & 120 & $5.4 \times 10^{-3}$ & 96 & $2.6 \times 10^{-3}$               \\
$(\frac{2}{5}, \frac{2}{5})$     &  (0,3)   &  $1 \times 2$       &  $3\frac{4}{5}$     & -3       & --                     & -- & --                     & --\\
$(\frac{7}{5}, \frac{7}{5})$     &  (0,1)   &  $1 \times 1$       & $3\frac{4}{5}$      & -1       & 72 & $8.3 \times 10^{-4}$   & 60 & $3.0 \times 10^{-3}$                  \\
$(\frac{2}{5}, \frac{7}{5})$      &  (1,1)   &  $1 \times 1$        &  $3\frac{4}{5}$    & -1         & --                  & -- & --                  & --\ \\
$(\frac{2}{5}, \frac{2}{5})$   &  (1,2)   &  $1 \times 2$           &  $3\frac{4}{5}$      & -1      & --                      & -- & --                  & --\\\
$(\frac{7}{5}, \frac{7}{5})$     &  (1,0)   &  $1 \times 1$       & $3\frac{4}{5}$      & +1   & 72 & $8.3 \times 10^{-4}$  & 60 & $3.0 \times 10^{-3}$                    \\
$(\frac{7}{5}, \frac{2}{5})$      &  (1,1)   &  $1 \times 1$        &  $3\frac{4}{5}$     & +1      & --                       & -- & --                  & --\\\
$(\frac{2}{5}, \frac{2}{5})$   &  (2,1)   &  $2 \times 1$           &  $3\frac{4}{5}$    & +1      & --                     & -- & --                  & --\\\
$(\frac{7}{5}, \frac{2}{5})$    &  (2,0)   &  $2 \times 1$          &  $3\frac{4}{5}$    & +3    & 120 & $5.4 \times 10^{-3}$  & 96 & $2.6 \times 10^{-3}$                    \\
$(\frac{2}{5}, \frac{2}{5})$   &  (3,0)   &  $2 \times 1$           &  $3\frac{4}{5}$    & +3      & --                     & --  & --                  & --\ \\
\hline
$(0,3)$                              &  (0,1)   &  $1 \times 1$          & 4                        & -4       & 120 & $7.7 \times 10^{-3}$  & 96 & $5.1 \times 10^{-3}$                   \\
$(0, 0)$                             &  (0,4)   &  $1 \times 2$           & 4                         & -4        & --                  & --  & --                  & -- \\
$(0, 0)$                               &  (2,2)   &  $1 \times 1$        & 4                 & 0         & 72 & $7.2 \times 10^{-4}$        & 60 & $1.7 \times 10^{-3}$           \\
$(3,0)$                              &   (0,0)   &  $2 \times 1$          & 4                        & +4     & 120 & $7.7 \times 10^{-3}$     & 96 & $5.1 \times 10^{-3}$                       \\
$(0, 0)$                              &  (4,0)   &  $1 \times 1$          & 4                         & +4       & --                     & --  & --                  & --  \\
\hline
$(\frac{2}{5}, \frac{7}{5})$   &  (0,3)  &  $1 \times 2$ &  $4\frac{4}{5}$  & -4 & 120 & $1.7 \times 10^{-2}$  & 96 & $1.7 \times 10^{-2}$               \\
$(\frac{2}{5}, \frac{2}{5})$   &  (0,4)  &  $1 \times 3$ &  $4\frac{4}{5}$  & -4 & --  & -- & --  & --\\
\hline
$(\frac{7}{5}, \frac{7}{5})$  &  (0,2)  &  $1 \times 2$  & $4\frac{4}{5}$ & -2 & 120 & $9.7 \times 10^{-3}$ & 94 & $6.0 \times 10^{-3}$                   \\
$(\frac{2}{5}, \frac{7}{5})$  &  (1,2)  &  $1 \times 2$  & $4\frac{4}{5}$ & -2 & -- & -- & --  & --\\
$(\frac{2}{5}, \frac{2}{5})$  &  (1,3)  &  $1 \times 2$  & $4\frac{4}{5}$ & -2 & -- & -- & --  & --\\
$(\frac{7}{5}, \frac{2}{5})$  &  (0,3)  &  $1 \times 2$  & $4\frac{4}{5}$ & -2 & -- & -- & --  & --\\
\hline
$(\frac{7}{5}, \frac{7}{5})$   &  (1,1)   &  $1 \times 1$  & $4\frac{4}{5}$  & +0   & 72 & $1.3 \times 10^{-3}$   & 60 & $1.9 \times 10^{-3}$                  \\
$(\frac{2}{5}, \frac{7}{5})$   &  (2,1)   &  $1 \times 1$  & $4\frac{4}{5}$  & +0   & --                       & -- & --  & --\\
$(\frac{2}{5}, \frac{2}{5})$   &  (2,2)   &  $1 \times 1$  & $4\frac{4}{5}$  & +0   & --                     & -- & --  & --\\
$(\frac{7}{5}, \frac{2}{5})$   &  (1,2)   &  $1 \times 1$  & $4\frac{4}{5}$  & +0   & --                       & -- & --  & --\\
\hline
$(\frac{7}{5}, \frac{7}{5})$   &  (2,0)   &  $2 \times 1$  & $4\frac{4}{5}$  & +2   & 120 & $9.7 \times 10^{-3}$     & 94 & $6.0 \times 10^{-3}$                  \\
$(\frac{7}{5}, \frac{2}{5})$   &  (2,1)   &  $2 \times 1$  & $4\frac{4}{5}$  & +2   & --  & --                     & --  & -- \\
$(\frac{2}{5}, \frac{2}{5})$   &  (3,1)   &  $2 \times 1$  & $4\frac{4}{5}$  & +2   & --  & -- & --  & --\\
$(\frac{2}{5}, \frac{7}{5})$   &  (3,0)   &  $2 \times 1$  & $4\frac{4}{5}$  & +2   & --  & -- & --  & --\\
\hline
$(\frac{7}{5}, \frac{2}{5})$   &  (3,0)   &  $2 \times 1$  & $4\frac{4}{5}$  & +4   & 120 & $1.7 \times 10^{-2}$  & 96 & $1.7 \times 10^{-2}$                  \\
$(\frac{2}{5}, \frac{2}{5})$   &  (4,0)   &  $3 \times 1$  & $4\frac{4}{5}$  & +4   & --  & -- & --  & --\\
\end{longtable}

\setlength{\LTleft}{0pt}
\setlength{\LTright}{0pt}
\begin{longtable}{ccccccccc}
\caption{\label{tab:3clock_q1} 3-state clock model, $Q=1$ sector} \\
\hline\hline
$\left(h,\bar{h}\right)$ & $\left( I, \bar{I} \right)$ & $d \times \bar{d}$ & $\Delta$ & $S$ & $L_x^*$ & $\text{RE}_\Delta (L_x^*)$ & $\widetilde{L}_x^*$ & $\text{RE}_\Delta (\widetilde{L}_x^*)$  \\
\hline\hline
\endfirsthead
\multicolumn{7}{c}{\tablename\ \thetable\ -- \textit{Continued}} \\
\hline\hline
$\left(h,\bar{h}\right)$ & $\left( I, \bar{I} \right)$ & $d \times \bar{d}$ & $\Delta$ & $S$ & $L_x^*$ & $\text{RE}_\Delta (L_x^*)$ & $\widetilde{L}_x^*$ & $\text{RE}_\Delta (\widetilde{L}_x^*)$  \\
\hline\hline
\endhead
\hline
\multicolumn{7}{r}{\textit{Continued on next page}} \\
\endfoot
\hline\hline
\endlastfoot
$(\frac{1}{15}, \frac{1}{15})$ & (0,0)    &  $1 \times 1$   & $\frac{2}{15}$     &  +0       & 144 & $1.4 \times 10^{-3}$   & 120 & $1.5 \times 10^{-3}$               \\
\hline
$(\frac{1}{15}, \frac{1}{15})$  & (0,1)    &  $1 \times 1$  & $1\frac{2}{15}$  &  -1   & 84 & $8.1 \times 10^{-4}$ & 72 & $7.8 \times 10^{-4}$                  \\
$(\frac{1}{15}, \frac{1}{15})$  & (1,0)    &  $1 \times 1$  & $1\frac{2}{15}$  &  +1   & 84 & $8.1 \times 10^{-4}$ & 72 & $7.8 \times 10^{-4}$                \\
\hline
$(\frac{2}{3}, \frac{2}{3})$      & (0,0)    &  $1 \times 1$  & $1\frac{1}{3}$    &  +0     & 96 & $4.3 \times 10^{-4}$ & 84 & $6.7 \times 10^{-4}$                      \\
\hline
$(\frac{1}{15}, \frac{1}{15})$  & (0,2)    &  $1 \times 2$  & $2\frac{2}{15}$  &  -1      & 120 & $5.7 \times 10^{-4}$ & 96 & $2.9 \times 10^{-4}$                    \\
$(\frac{1}{15}, \frac{1}{15})$  & (1,1)    &  $1 \times 1$  & $2\frac{2}{15}$  &  +0     & 60 & $2.8 \times 10^{-3}$    & 48 & $2.8 \times 10^{-3}$                  \\
$(\frac{1}{15}, \frac{1}{15})$  & (2,0)    &  $2 \times 1$ & $2\frac{2}{15}$  &  +1      & 120 & $5.7 \times 10^{-4}$  & 96 & $2.9 \times 10^{-4}$                   \\
\hline
$(\frac{2}{3}, \frac{2}{3})$      & (0,1)    &  $1 \times 1$  & $2\frac{1}{3}$     & -1         & 96 & $1.3 \times 10^{-4}$ & 84 & $7.9 \times 10^{-4}$                    \\
$(\frac{2}{3}, \frac{2}{3})$      & (1,0)    &  $1 \times 1$  & $2\frac{1}{3}$    & +1          & 96 & $1.3 \times 10^{-4}$ & 84 & $7.9 \times 10^{-4}$                   \\
\hline
$(\frac{1}{15}, \frac{1}{15})$  & (0,3)    &  $1 \times 3$  & $3\frac{2}{15}$  &  -2    & 120 & $3.7 \times 10^{-3}$  & 96 & $2.1 \times 10^{-3}$                        \\
$(\frac{1}{15}, \frac{1}{15})$  & (1,2)    &  $1 \times 2$  & $3\frac{2}{15}$  &  -1     & 84 & $1.4 \times 10^{-3}$  & 72 & $2.5 \times 10^{-3}$                   \\
$(\frac{1}{15}, \frac{1}{15})$  & (2,1)    &  $2 \times 1$  & $3\frac{2}{15}$   &  +1    & 84 & $1.4 \times 10^{-3}$   & 72 & $2.5 \times 10^{-3}$                   \\
$(\frac{1}{15}, \frac{1}{15})$  & (3,0)    &  $3 \times 1$ & $3\frac{2}{15}$   &  +2    & 120 & $3.7 \times 10^{-3}$    & 96 & $2.1 \times 10^{-3}$                      \\
\hline
$(\frac{2}{3}, \frac{2}{3})$      & (0,2)    &  $1 \times 2$  & $3\frac{1}{3}$     & -2     & 120 & $3.7 \times 10^{-3}$ & 96 & $1.4 \times 10^{-3}$                      \\
$(\frac{2}{3}, \frac{2}{3})$      & (1,1)    &  $1 \times 1$  & $3\frac{1}{3}$     & 0      & 84 & $6.5 \times 10^{-6}$   & 72 & $1.3 \times 10^{-3}$                    \\
$(\frac{2}{3}, \frac{2}{3})$      & (2,0)    &  $2 \times 1$  & $3\frac{1}{3}$     & +2   & 120 & $3.7 \times 10^{-3}$  & 96 & $1.4 \times 10^{-3}$                      \\
\hline
$(\frac{1}{15}, \frac{1}{15})$  & (0,4)    &  $1 \times 5$  & $4\frac{2}{15}$   &  -4      & 120 & $9.4 \times 10^{-3}$  & 96 & $6.9 \times 10^{-3}$                 \\
$(\frac{1}{15}, \frac{1}{15})$  & (1,3)    &  $1 \times 3$  & $4\frac{2}{15}$   &  -2         & 84 & $1.8 \times 10^{-3}$ & 72 & $1.4 \times 10^{-4}$                \\
$(\frac{1}{15}, \frac{1}{15})$  & (2,2)    &  $2 \times 2$  & $4\frac{2}{15}$  &  +0       & 72 & $2.3 \times 10^{-4}$  & 60 & $2.0 \times 10^{-3}$                        \\
$(\frac{1}{15}, \frac{1}{15})$  & (3,1)    &  $3 \times 1$  & $4\frac{2}{15}$   &  +2       & 84 & $1.8 \times 10^{-3}$  & 72 & $1.4 \times 10^{-4}$                 \\
$(\frac{1}{15}, \frac{1}{15})$  & (4,0)    &  $5 \times 1$ & $4\frac{2}{15}$   &  +4    & 120 & $9.4 \times 10^{-3}$   & 96 & $6.9 \times 10^{-3}$                  \\
\hline
$(\frac{2}{3}, \frac{2}{3})$      & (0,3)    &  $1 \times 2$  & $4\frac{1}{3}$    & -3        & 120 & $1.2 \times 10^{-2}$  & 96 & $8.2 \times 10^{-3}$                       \\
$(\frac{2}{3}, \frac{2}{3})$      & (1,2)    &  $1 \times 2$  & $4\frac{1}{3}$    & -1        & 96 & $3.6 \times 10^{-3}$   & 84 & $1.7 \times 10^{-3}$                     \\
$(\frac{2}{3}, \frac{2}{3})$      & (2,1)    &  $2 \times 1$  & $4\frac{1}{3}$    & +1        & 96 & $3.6 \times 10^{-3}$  & 84 & $1.7 \times 10^{-3}$                      \\
$(\frac{2}{3}, \frac{2}{3})$      & (3,0)    &  $2 \times 1$  & $4\frac{1}{3}$    & +3       & 120 & $1.2 \times 10^{-2}$   & 96 & $8.2 \times 10^{-3}$                      \\
\end{longtable}

\begin{acknowledgments}
  We thank Wenhan Guo, Tzu-Chieh Wei, Atsushi Ueda for helpful discussions.
  We also acknowledge the use of AI-based language tools for editing and polishing the manuscript.
  This work was supported by the National Science and Technology Council (NSTC) of Taiwan through Grant Nos. 112-2119-M-007-008 and 113-2119-M-007-008.
\end{acknowledgments}

\bibliography{references}

\end{document}